\documentclass[floatfix,reprint,amsmath,amssymb,superscriptaddress,aps,prl]{revtex4-2}
\usepackage[american]{babel}
\usepackage{graphicx}
\usepackage{bm}
\usepackage[hidelinks]{hyperref}
\usepackage[capitalize]{cleveref}
\usepackage[nice]{nicefrac}
\usepackage{physics}
\usepackage{xcolor}
\usepackage{xspace}
\usepackage{mathtools}  
\usepackage[normalem]{ulem} 
\usepackage{hyperref}
\usepackage{xcolor}
\hypersetup{colorlinks,bookmarksopen,bookmarksnumbered,
citecolor=teal,
linkcolor=teal,
pdfstartview=false,
urlcolor=teal}

\begin{document}
\date{\today}

\title{Dissipative phase transition of interacting non-reciprocal fermions}

\author{Rafael D. Soares}
\affiliation{Max Planck Institute for the Physics of Complex Systems, N\"{o}thnitzer Stra{\ss}e 38, 01187 Dresden, Germany}
\affiliation{JEIP, UAR 3573 CNRS, Coll\`ege de France, PSL Research University, 11 Place Marcelin Berthelot, 75321 Paris Cedex 05, France}
\author{Matteo Brunelli}
\affiliation{JEIP, UAR 3573 CNRS, Coll\`ege de France, PSL Research University, 11 Place Marcelin Berthelot, 75321 Paris Cedex 05, France}
\author{Marco Schir\`o}
\affiliation{JEIP, UAR 3573 CNRS, Coll\`ege de France, PSL Research University, 11 Place Marcelin Berthelot, 75321 Paris Cedex 05, France}

\begin{abstract}
While non-reciprocal couplings are ubiquitous in classical systems, their impact on quantum many-body criticality and entanglement remains largely unexplored. Using exact numerical simulations, we study an interacting fermionic chain subject to non-reciprocal gain and loss. We show that the interplay between dissipation and interactions drives a dissipative phase transition, marked by the opening of a many-body gap and a crossover from power-law to exponential relaxation. The weakly-interacting regime displays non-reciprocal signatures, including nonzero currents and directional charge accumulation reminiscent of the skin effect. Notably, despite this localization, quantum trajectories exhibit volume-law entanglement. Finally, reciprocity is dynamically restored above a critical interaction strength.
\end{abstract}

\maketitle
    
\emph{Introduction -- } For classical and quantum systems in thermal equilibrium, the interactions between degrees of freedom are fundamentally symmetric or reciprocal. Far from equilibrium, however, asymmetric or non-reciprocal interactions are ubiquitous. Examples range from optics~\cite{caloz2018whatis,sounas2017nonreciprocal}, to active~\cite{zhihong2020nonreciprocity,saha2020scalar,poncet2022when,brauns2024nonreciprocal} or living matter~\cite{Tan_2022_livingmatter}, ecology~\cite{ros2023generalized,blumenthal2024phase}, robotic systems~\cite{Brandenbourger_2019_robotics} and statistical physics~\cite{lorenzana2024nonreciprocalspinglasstransitionaging,hanai2024nonreciprocal,avni2025nonreciprocal,young2024nonequilibriumuniversalitynonreciprocallycoupled}. Non-reciprocal systems display unusual properties, including unique critical behavior at phase transitions~\cite{fruchart2021nonreciprocal}. First investigated for classical systems, interest in non-reciprocity has branched out to the quantum world. Non-reciprocal phenomena can naturally arise in non-Hermitian (NH) quantum mechanics~\cite{ashida2020review}, the Hatano-Nelson model providing a well-known example of a system with  non-reciprocal hopping~\cite{hatano1996localization,hatano1997vortex}, leading to rich physics such as enhanced sensitivity to boundary conditions, the so-called NH skin effect~\cite{lee2016anomalous,yao2018edge,sato2023nonhermitian}, NH topology~\cite{Gong_2018_nhtopology, kawabata2019symmetry} and entanglement phase transitions~\cite{kawabata2023entanglement}. However, this approach neglects the contribution from quantum jumps. A fully consistent quantum-mechanical treatment of non-reciprocity is possible within the framework of cascaded master equations~\cite{Gardiner_1993, carmichael1993quantum} and reservoir engineering~\cite{metelman2015nonreciprocal,clerk2022introduction}.

The first steps towards applying these ideas to open quantum many-body systems~\cite{fazio2025manybodyopenquantumsystems} have recently been taken~\cite{keck2018persistent,hanai2019nonhermitian,Song2019ChiralDamping,yang2022liouvillian,McDonald2022NonEq,chiacchio2023nonreciprocal,Begg2024,brighi2024nonreciprocal,nadolny2024nonreciprocalsynchronizationactivequantum, Nadolny_2025_laser, zelle2024universal,belyansky2025phasetransitionsnonreciprocaldrivendissipative}. Outstanding questions are whether quantum many-body systems can exhibit robust non-reciprocal steady states featuring rich entanglement structures, akin to those found in their NH counterparts, and whether quantum non-reciprocal phase transitions connecting such states can exist. In addition, one could ask whether these phase transitions can be classified within the framework of dissipative phase transitions or whether they instead represent a new universality class~\cite{sieberer2023universalitydrivenopenquantum}.

\begin{figure}[t]
\centering
\includegraphics[width=0.9\linewidth]{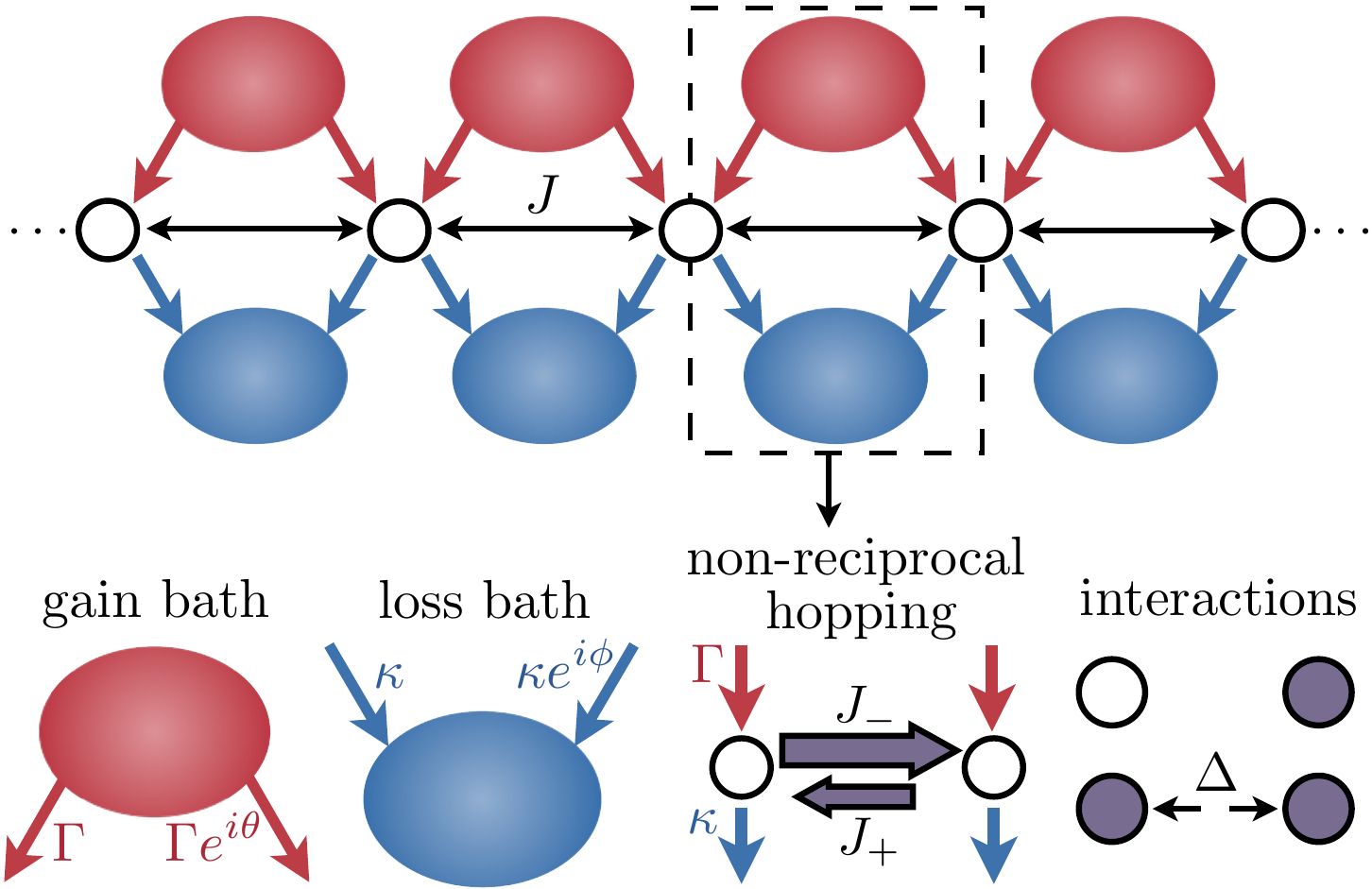}
\caption{Sketch of the setup: one-dimensional fermionic chain with nearest-neighbor hoppings, $J$, and density-density interaction, $\Delta$. The gain (loss) bath injects (absorbs) particles to (from) neighboring sites, with relative phase $\theta$ ($\phi$). Together with the hopping term, they realize non-reciprocal gain and loss processes.}
\label{fig:sketch}
\end{figure}

In this work, we consider an interacting fermionic chain featuring non-reciprocal hopping engineered via active and passive reservoirs (Fig.~\ref{fig:sketch}). The resulting gain and loss rates are characterized by non-trivial relative phases yielding an asymmetric coupling matrix---processes we refer to as non-reciprocal gain and loss. Crucially, these mechanisms can balance each other to produce gapless, critical non-reciprocal dynamics in the non-interacting limit. This contrasts with purely local gain~\cite{Begg2024}, which always exhibits a finite dissipative gap. Using large-scale diagonalization of the Lindblad master equation and quantum trajectories, we reveal an interaction-induced transition marked by the opening of a finite dissipative gap and the onset of exponential relaxation. At this transition, reciprocity is dynamically restored, as evidenced by the suppression of boundary charge accumulation. Remarkably, unlike effective NH Hamiltonians featuring NH skin effect~\cite{kawabata2023entanglement}, the weakly interacting phase supports volume-law trajectory entanglement despite exhibiting identical boundary charge accumulation. Our work highlights the role of non-reciprocal gain and loss in giving rise to exotic dynamical and steady-state properties.

\emph{Model and Setup --} We consider an interacting fermionic chain of size $L$ described by the Hamiltonian
\begin{equation}\label{eqn:H}
\mathcal{H}=-\dfrac{J}{2}\sum_{j=1}^{L-1}\left[c_{j}^{\dagger}c_{j+1}+c_{j+1}^{\dagger}c_{j}\right]+\Delta\sum_{j=1}^{L-1}n_jn_{j+1},
\end{equation}
where $c_{j}^{\dagger}\left(c_{j}\right)$ is the fermionic creation(annihilation) operator that creates(annihilates) a particle at site $j$, $n_j=c_{j}^{\dagger}c_{j}$, $J$ is the hopping strength and $\Delta$ is the next-neighbor density-density interaction, see Fig.~\ref{fig:sketch}. We are interested in the dissipative dynamics generated by the Linblad master equation for the system density matrix $\rho$~\cite{fazio2025manybodyopenquantumsystems}
\begin{equation}\label{eqn:lindblad}
\partial_{t}\rho=-i\left[\mathcal{H},\rho\right]+\sum_{j,\mu}L_{j,\mu}\rho L_{j,\mu}^{\dagger}-\dfrac{1}{2}\left\{ L_{j,\mu}^{\dagger}L_{j,\mu},\rho\right\} ,
\end{equation}
where $L_{j,\mu}$ are a set of jump operators describing non-local gain and losses ($\mu=g,\ell$ respectively) of strength $\Gamma,\kappa$, that we take of the form 
\begin{equation}
\label{eqn:jumps}
L_{j,\ell}=\sqrt{\kappa}\left(c_{j}+e^{i\phi}c_{j+1}\right),\,L_{j,g}=\sqrt{\Gamma}\left(c_{j}^{\dagger}+e^{i\theta}c_{j+1}^{\dagger}\right).
\end{equation}
A key feature of these jump operators is the relative phases $\phi,\theta$ between the rates of loss and gain on neighboring sites. These phases can be seen to arise from the coupling to environments featuring synthetic gauge fields. Jump operators of this form can be realized within the framework of reservoir engineering~\cite{metelman2015nonreciprocal, Malz_2018_reseng, McDonald2022NonEq, Wanjura_2020_topamp,Porras_2019_amp, Brunelli_2023_correspondence}, in various platforms including superconducting circuits and trapped ions~\cite{doi:10.1126/sciadv.adj8796,PhysRevApplied.22.034038,PhysRevX.13.021021}. For $\theta,\phi\neq 0$ the jump operators in Eq.~(\ref{eqn:jumps}) give rise to asymmetric gain and loss matrices, i.e., to non-reciprocal dissipative couplings. In the absence of coherent hopping ($J=0$), one of the two phases can be removed from the Lindblad master equation by a local gauge transformation~\cite{supplementary_mat}. On the other hand, for $J\neq0$, the two phases give rise to a rich landscape of non-reciprocal behavior in dynamic and static properties, as we  show below.
\begin{figure*}[t]
    \centering
    \includegraphics{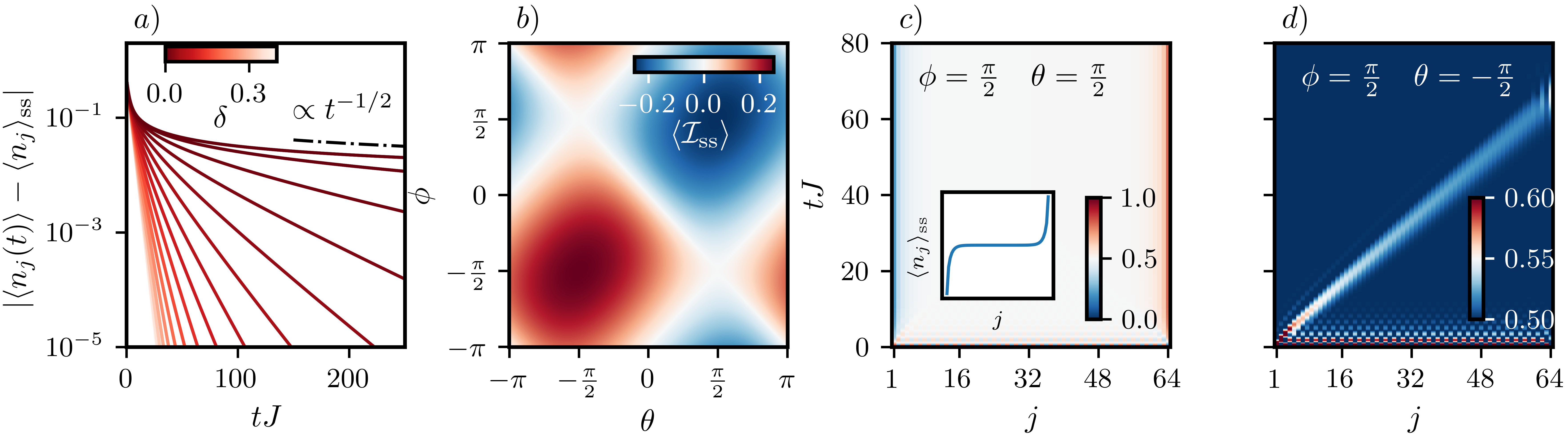}
\caption{ $a)$ - Time evolution of the particle density under PBC (rescaled by the steady-state value) for different values of the dissipative gap $\delta$. $b)$ - Steady-state particle current under PBC as a function of the angles. In $c)$ and $d)$, we depict the particle density dynamics with OBC for different values of the angles, with the initial state a CDW. The inset in $c)$ corresponds to the steady-state configuration. Other parameters: $\Gamma=\kappa=0.1J$.}
\label{fig:non_interacting}
\end{figure*}

\emph{Non-Interacting Limit --} We start by discussing the non-interacting limit $\Delta=0$, which provides useful insight (see also Ref.~\cite{Song2019ChiralDamping} where the case $\phi=-\theta=\pi/2$ has been studied for a Su–Schrieffer–Heeger model).

The dynamics of the correlation matrix $C_{ij}=\langle c^{\dagger}_i c_j\rangle $ takes the form
\begin{equation}\label{eq:uncond}
i\partial_t C_{ij}=\sum_{l}\left(H\right)_{jl}C_{il}-
\left(H^\dagger\right)_{li}C_{lj}+iG_{ij},
\end{equation}
where $\left(H\right)_{jl}=-i\left(\kappa+\Gamma\right)\delta_{jl}-\frac{1}{2}J_{+}\delta_{j,l+1}-\frac{1}{2}J_{-}\delta_{j,l-1}$ encodes an effective NH Hamiltonian associated with the unconditional evolution, which in the presence of gain is different from the no-click limit~\cite{McDonald2022NonEq}, and noise matrix $G_{jl}=\Gamma \left(e^{-i\theta}\delta_{j,l+1}+e^{i\theta}\delta_{j,l-1} + 2\delta_{j,l}\right)$. In our case $H$ takes the form of the Hatano-Nelson model~\cite{hatano1996localization} with hopping amplitudes 
\begin{equation}\label{eqn:Jpm}
J_{\pm}=J+i\left(\Gamma\cos\theta+\kappa\cos\phi\right)\pm\left(\Gamma\sin\theta-\kappa\sin\phi\right).
\end{equation}
Unequal amplitudes $J_+ \neq J_-$ signify non-reciprocal hopping.

For periodic boundary conditions (PBC), we can compute exactly the  dynamics of the occupation numbers~\cite{supplementary_mat} $\langle n_k \rangle =\langle c^\dagger_k c_k \rangle$, which read 
\begin{equation}\label{eqn:nk_t}
\langle n_{k} (t) \rangle = \langle n_{k} \rangle_{\text{ss}} + e^{-\lambda_k t} \left(\langle n_{k} (0) \rangle - \langle n_{k} \rangle_{\text{ss}} \right),
\end{equation}
where $\lambda_k$ represents the spectrum of decay rates
\begin{equation}\label{eqn:spectrum}
\lambda_k=2(\Gamma+\kappa)+\mbox{Im}(J_++J_-)\cos k+
(J_+-J_-)\sin k
\end{equation}
while $\langle n_{k}\rangle_{\text{ss}}$ is the steady-state distribution
\begin{equation}\label{eqn:nkss}
\langle n_{k}\rangle_{\text{ss}} = \frac{\Gamma(1+\cos\theta\cos k+\sin\theta\sin k)}{2(\Gamma+\kappa)+\mbox{Im}(J_++J_-)\cos k+
(J_+-J_-)\sin k}\,.
\end{equation}
We note that the phases of gain and losses $\theta,\phi$ enter differently in the spectrum $\lambda_k$ and in the steady-state occupation $\langle n_{k}\rangle_{\text{ss}}$. As a result, the two dissipative processes give rise to distinct non-reciprocal behavior in the dynamics or in the stationary state.

We start by discussing the spectrum and the dynamics. From Eq.~(\ref{eqn:spectrum}), we see that the slowest decaying mode is at $k_*=\arctan[\left(\Gamma\sin\theta-\kappa\sin\phi\right)/\left(\Gamma\cos\theta+\kappa\cos\phi\right)]$ and around this value the spectrum reads $\lambda_k=\delta+\alpha(k-k_*)^2$, with the dissipative gap
\begin{equation}
    \delta = 2 \left(\Gamma +\kappa\right)-2\sqrt{\Gamma^2+\kappa^2+2\Gamma\kappa\cos\left(\theta+\phi\right)}\;.
\end{equation}
In the $\phi,\theta$ plane the gap is finite everywhere, except along the line $\phi=-\theta$ where it closes as $\delta\propto|\phi-\phi_c|^2$. This implies that an excitation with wave-vector $k_*$ is not affected by dissipation and evolves due to the unitary part of evolution alone~\cite{supplementary_mat}. The existence of this gapless mode has direct consequences on the system's dynamics, making observables, such as the particle density, approach the steady state following a universal power-law decay~\cite{supplementary_mat} $\langle n(t)\rangle\sim \langle n\rangle_{\text{ss}}-\delta n/\sqrt{t}$. Moving away from this line the dissipative gap opens and the dynamics becomes exponential; see Fig.~\ref{fig:non_interacting}(a). This corresponds to a dissipative transition where the time scale to reach the steady-state diverges. Importantly, the closure of the dissipative gap along the critical line $\phi=-\theta$ arises from the interference between non-reciprocal gain and loss. It therefore stands in contrast to the case of purely local gain~\cite{Begg2024}, where the system is always gapped at finite dissipation, irrespective of the reciprocal vs non-reciprocal character of the dynamics (i.e. for any value of $\phi$). In our system, in the presence of non-reciprocal losses, non-reciprocal gain is necessary for closing the gap. The dynamics along the line $\phi=-\theta$ shows both critical and non-reciprocal features, since the spectrum is asymmetric in momentum space $\lambda_k-\lambda_{-k}= 2(J_+-J_-)\sin k$, leading to a non-reciprocal transient occupation $\langle n_{-k}(t)\rangle\neq \langle n_k(t)\rangle$ and so to a non-zero transient current $\langle \mathcal{I}\rangle\sim \mathcal{I}_0\sin\phi/\sqrt{t}$ that slowly vanishes at long times with a magnitude controlled by the phase of the gain and loss baths~\cite{supplementary_mat}.




We now discuss the steady-state distribution Eq.(\ref{eqn:nkss}), which features two distinct sources of non-reciprocity, namely from the spectrum (denominator) and from the 
non-reciprocal gain (numerator). This leads to a non-trivial distribution across the $\phi,\theta$ plane. Along the line $\phi = \theta$ and for $\Gamma=\kappa$, where the hopping is reciprocal ($J_+ = J_-$), the steady-state distribution in Eq.~(\ref{eqn:nkss}) is highly non-thermal~\cite{supplementary_mat}. Furthermore, the presence of non-reciprocal gain breaks inversion symmetry, as reflected in $\langle n_{-k} \rangle_{\text{ss}}\neq \langle n_{k}\rangle_{\text{ss}}$. This asymmetry results in a finite steady-state current, which reaches its maximum value at $\phi = \theta = \pi/2$~\cite{supplementary_mat}. On the other hand, along the line $\phi=-\theta$, the two dissipative processes interfere destructively and the distribution becomes flat in momentum space, $\langle n_{k} \rangle_{\text{ss}}=\Gamma/(\Gamma+\kappa)$, and as a consequence the steady-state current vanishes, just as in the case of local bulk gain and losses. 
When plotted in the $\phi,\theta$ plane, the steady-state current shows a diamond structure reminiscent of Coulomb blockade physics~\cite{Nazarov:1186214}, with lines of zero and maximal currents that arise due to interference between the dissipative processes, see Fig.~\ref{fig:non_interacting}(b).

The steady-state correlation matrix decays exponentially across the phase diagram with a characteristic length that grows approaching the line $\phi=-\theta$~\cite{supplementary_mat}. However, when the dissipative gap closes also the momentum dependent gain vanishes, preventing the correlation length to diverge~\cite{zhang2022criticality,Barthel_2022}. Notably, along this line the system relaxes to a maximally mixed state with a particle density fixed by the ratio of $\Gamma/\kappa$~\cite{supplementary_mat}.

We now consider the dynamics with open boundary conditions (OBC)\footnote{To ensure that all sites are subject to the same overall particle gain and losses, we add an extra loss/gain processes at the boundaries, described by jump operators $L_{0,\ell}=\sqrt{\kappa}c_{0},$ and $L_{L-1,\ell}=\sqrt{\kappa}c_{L-1}$, similarly for the pump $L_{0,g}=\sqrt{\Gamma}c_{0}^{\dagger}$ and $L_{L-1, g}=\sqrt{\Gamma}c_{L-1}^{\dagger}$.}. In Fig.~\ref{fig:non_interacting}(c-d), we plot the dynamics of the particle density for different values of $\theta,\phi$ and different initial states, at $\Gamma=\kappa$. In stark contrast to the PBC case, the relaxation towards the steady-state is strictly exponential as the dissipative gap remains always finite, a manifestation of the Liouville skin-effect~\cite{supplementary_mat}. For $\phi=\theta=\pi/2$, see Fig.~\ref{fig:non_interacting} (c), the initial charge-density wave (CDW) is quickly melted in the bulk after a few short-time revivals, in agreement with the previous discussion of exponential dynamics. However, at the edges the dynamics is nontrivial: charge is reshuffled across the system by the non-reciprocal gain and loss, and we observe the build up of a steady state featuring charge accumulation at the edges.  Upon tuning the angle $\theta$ we see that the charge imbalance decreases and eventually disappears for $\theta=-\pi/2$, as the steady-state corresponds to the maximally mixed state~\cite{supplementary_mat}, a value at which, however, the dynamics displays clear directionality; see Fig.~\ref{fig:non_interacting}.(d).
We can understand these results by looking at the non-reciprocal hopping rates $J_\pm$ in 
Eq.~(\ref{eqn:Jpm}). For $\Gamma=\kappa$ the maximum value of $J_+-J_-$ is obtained precisely for $\phi=-\theta=\pi/2$, while for $\phi=\theta=\pi/2$ the hopping is perfectly reciprocal and so the dynamics. Overall, this behavior is reminiscent of the dynamics of the Hatano-Nelson model, yet it appears at the level of a fully Lindbladian evolution Eq.~\eqref{eq:uncond}, which includes dissipation and noise and does not require post-selection~\footnote{We note that, on the other hand, the fact that $J_++J_-$ is complex does not induce any non-reciprocal or unidirectional dynamics since the imaginary part of the Hamiltonian is proportional to the kinetic energy and not the current.}.

\begin{figure}[t]
    \centering
    \includegraphics{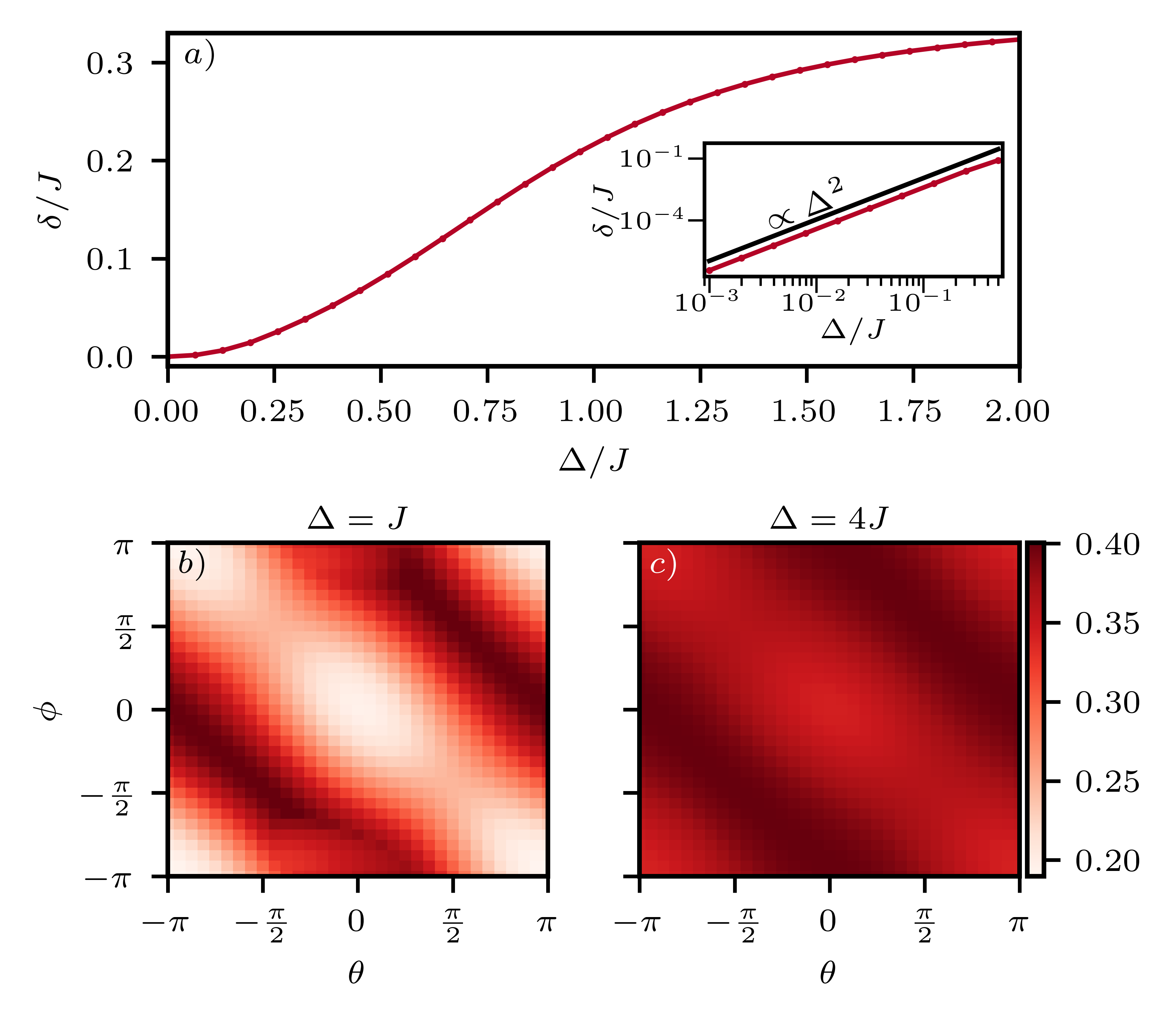}
    \caption{Dissipative gap for PBC. $a)$ Gap as a function of the interaction strength ($\phi = -\theta = \pi/2, L = 16$). $b), c)$ Gap as a function of $\phi$ and $\theta$ for $L = 12$ with $\Delta = J$ [$b)$] and $\Delta = 4J$ [$c)$]. Other parameters: $\Gamma = \kappa = 0.1J$.}
    \label{fig:dissipative_gap}
\end{figure}

\emph{Interaction-induced dissipative transition--} Turning to the interacting case, we perform large-scale exact diagonalization of the many-body Lindbladian to investigate the dissipative gap $\delta$~\cite{supplementary_mat}. We first focus on the non-interacting critical line $\phi=-\theta=\pi/2$, where the gap either vanishes exactly (if the system resolves the dark mode momentum $k_*$) or closes in the thermodynamic limit as $\delta\sim 1/L^2$~\cite{supplementary_mat}. Upon adding interactions, Fig.~\ref{fig:dissipative_gap}(a) shows that $\delta$ scales quadratically for weak coupling before increasing rapidly near $\Delta\sim J$. While the gap remains finite in finite-size systems, this suggests a rapid crossover at weak interactions. Tuning the angles [Fig.~\ref{fig:dissipative_gap}(b)] reveals that up to $\Delta\sim J$, the gap retains a structure similar to the non-interacting case, with a minimum along $\phi=-\theta$. This dependence persists at intermediate $\Delta$ but is suppressed as interactions increase, approaching a uniform gap [Fig.~\ref{fig:dissipative_gap}(c)].

In addition to the behavior of the dissipative gap, our numerically exact results give access to the symmetry-resolved many-body spectrum based on translation invariance and weak $U(1)$ symmetry, thus providing insight into the structure of the slowest mode~\cite{supplementary_mat}. In the non-interacting case, along the critical line, we found four eigenvalues with zero real-part corresponding to two degenerate steady states with zero momentum and two states featuring single particle excitations at $k_*$. While a finite interaction partially lifts these degeneracies, there is a many-body level crossing separating the weak-interaction regime, where the slowest mode is at finite momentum, from the strongly interacting regime, where it belongs to the zero momentum sector and connects to the non-interacting steady-state as $\Delta\rightarrow0$. These results suggest that our system for $\phi=-\theta$ undergoes a dissipative phase transition driven by interactions via the opening of a many-body gap.

\begin{figure}[b]
    \centering
    \includegraphics{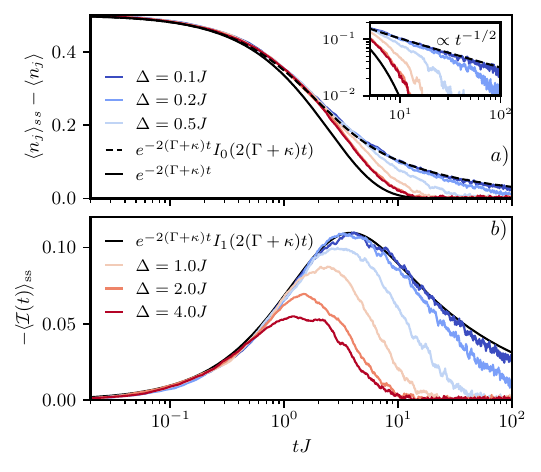}
    \caption{Time evolution of $a)$ the particle density (rescaled by the steady-state value) and $b)$ the particle current for different values of the interaction strength $\Delta$ with PBC for $\phi=-\theta=\pi/2$; the system is initialized in the empty state. $I_\alpha(\cdot)$ correspond to the $\alpha$ order Bessel function of the second kind~\cite{supplementary_mat}. Other parameters: $L=16$, $\Gamma=\kappa=0.1J$.}
    \label{fig:pbc_interacting_dynamics}
\end{figure}
To complement the picture obtained by the dissipative spectrum, we focus on dynamics and solve the problem by unraveling Eq.~\eqref{eqn:lindblad} into quantum jump trajectories~\cite{dalibard1992wavefunction,plenio1998quantum,daley2014quantum,fazio2025manybodyopenquantumsystems}, where the NH evolution between subsequent quantum jumps is made with the recently developed Faber polynomial method~\cite{supplementary_mat,Soares2024Faber}. We consider a finite chain with $\Gamma=\kappa=0.1J$ and PBC, fixing $\phi=-\theta=\pi/2$ to test whether the power-law relaxation survives for finite interactions. In Fig.~\ref{fig:pbc_interacting_dynamics}(a) we plot the dynamics of particle density for different values of $\Delta$, while negligible at short times, interaction effects dominate the long-time decay towards the steady state.

\begin{figure}[t]
    \centering
    \includegraphics{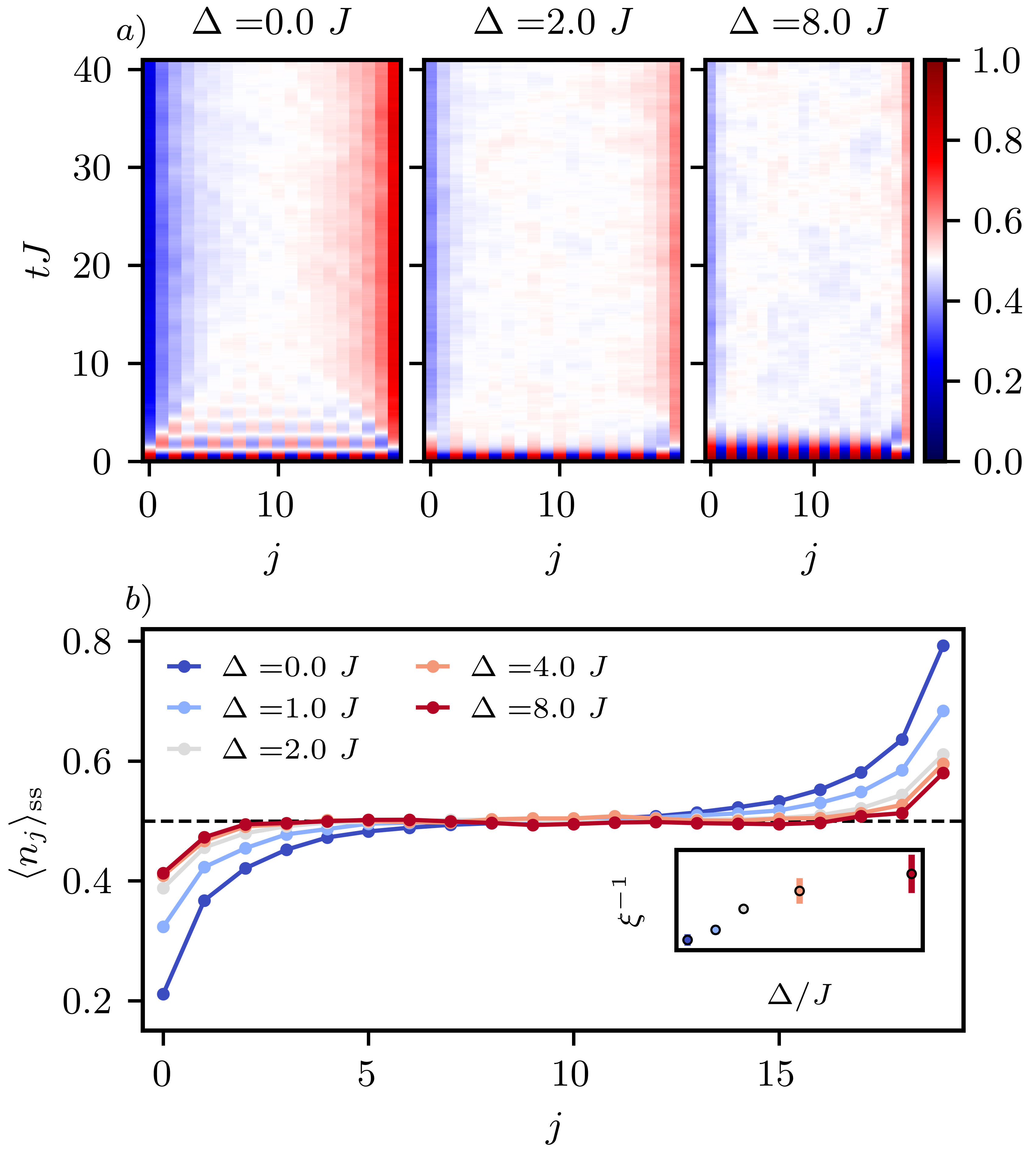}
    \caption{$a)$ Time evolution of the particle density and $b)$ its value in the steady state for different values of $\Delta$ and with OBC and $\phi =\theta= \pi/2$. The system is initially in a CDW. Other parameters: $\Gamma = \kappa = 0.1J$.}
    \label{fig:dynamics_interaction}
\end{figure}

Remarkably, for weak $\Delta$, the dynamics remains compatible with a power-law (the dashed line corresponds to the non-interacting case), possibly with an interaction-dependent exponent. However, for $\Delta\sim J$, the particle density exhibits a crossover to exponential decay governed by the total dissipation rate $2(\Gamma+\kappa)$. This points towards a genuine dissipative phase transition arising from the competition between interactions and non-reciprocity. Similar behavior is observed for the current (Fig.~\ref{fig:dynamics_interaction}(b)) which vanishes asymptotically since the steady state remains maximally mixed for $\phi=-\theta$ independently of $\Delta$~\cite{supplementary_mat}. At intermediate times, the peak in the time-dependent current is renormalized by interactions and pushed to shorter time scales. At long times, for $\Delta<J$, the decay is compatible with a power-law, but is rapidly suppressed at larger $\Delta$ .

\emph{Dynamics and Steady-State under OBC --} We now discuss the role of interactions in the OBC case. We consider the dynamics of the average particle density starting from a CDW state and for different values of $\Delta$. We fix $\phi=\theta=\pi/2$, where in the non-interacting case there is charge accumulation at the boundaries, see Fig.~\ref{fig:non_interacting}(c). As we show in Fig.~\ref{fig:dynamics_interaction}(a) increasing $\Delta$ has two major effects: (i) the tendency toward charge accumulation at the boundaries is reduced, although not suppressed completely even for $\Delta=8J$, and (ii) the initial CDW state remains frozen for longer times. These two effects are indeed connected and result from an interaction-induced renormalization of non-reciprocal gain and loss.

To understand this, we derive an effective model in the regime $\Delta\gg J,\Gamma,\kappa$. Moving to a rotating frame yields modified jump operators (analogous for $L_{j,g}$)~\cite{supplementary_mat}
\begin{align}\label{eqn:TildeL}
\Tilde{L}_{j,\ell }=\sqrt{\kappa}(e^{i\Delta (n_{j-1}+n_{j+1})t} c_j+ e^{i\phi+i\Delta (n_{j}+n_{j+2})t} c_{j+1}).
\end{align}
In this rotating frame, the Lindbladian is time-periodic with terms rapidly oscillating at frequency $\omega\sim \Delta$, which can be averaged out as in the high-frequency expansion of Floquet systems~\cite{bukov2016schrieffer,seetharam2018absence,peronaci2018resonant}. From Eq.~(\ref{eqn:TildeL}), we see that dissipative processes involving occupied neighbors are highly off-resonant and effectively blocked. This simple argument explains why in the strongly interacting regime the initial CDW state, containing no doublons, remains long-lived and the non-reciprocal gain and loss become less efficient, reducing the charge accumulation at the edge. It is interesting to characterize in more detail the structure of the steady-state density profile, which we plot in Fig.~\ref{fig:dynamics_interaction}(b) for different values of $\Delta$ with $\kappa=\Gamma=0.1 J$ and $\phi=\theta=\pi/2$. We see that the charge density in the bulk is pinned to half-filling, because of equal gain and loss rates. The left(right) boundary displays charge depletion(accumulation), with a decay towards the bulk that is compatible with an exponential law $\left|\langle n_j\rangle_{\text{ss}}-\langle n_{L/2}\rangle _{\text{ss}}\right|=1/2+Ae^{-j/\xi}$ where $\xi$ is the localization length and $A$ a fitting constant. In the inset, we show that the $\xi^{-1}$ increases with $\Delta$ and appears to saturate for $\Delta/J\gg 1$. Overall, these results suggest that interactions compete with the non-reciprocal gain and loss to reduce the tendency toward charge accumulation. Quite interestingly, in this regime, although the charge is localized at the edges and the skin effect would lead to area-law entanglement of NH free fermions~\cite{kawabata2023entanglement}, the unraveled steady-state for the interacting problem shows volume-law entanglement (see End Matter for details). To conclude, our numerical results support the emergence of a robust non-reciprocal interacting many-body state at weak couplings.

\emph{Conclusions --} We investigated the dynamics of an interacting fermionic chain subject to non-reciprocal gain and loss, in which non-reciprocity is controlled via two independent phases $\phi$, $\theta$. Using exact numerical simulations, we demonstrated the existence of a dissipative phase transition driven by the interplay between non-reciprocity and interactions. In the non-interacting limit, the single-particle gap closes along the critical line $\phi=-\theta$.  Crucially, this shows that in order for criticality to persist at finite dissipation rates, non-reciprocal loss ($\phi \neq 0$) must be accompanied by non-reciprocal gain ($\theta \neq 0$).
We then show that adding interactions opens a many-body gap along this critical line, inducing a crossover from power-law to exponential relaxation. While the steady state on the critical line remains maximally mixed, deviations from it reveal non-reciprocal signatures, including currents due to broken inversion symmetry.  Notably, despite a density profile reminiscent of the skin effect, unraveling the interacting steady-state into quantum trajectories shows a volume-law entanglement entropy. Our work highlights the rich transport and entanglement properties emerging from the competition between non-reciprocal dynamics and many-body interactions.

\emph{Acknowledgments --} We acknowledge A. Clerk, G. Lee and A. Pocklington for inspiring discussions. M.B and M.S. acknowledge funding from the European Research Council (ERC) under the European Union's Horizon 2020 research and innovation program (Grant agreement No. 101002955 -- CONQUER). We acknowledge Coll\`{e}ge de France IPH cluster where the numerical calculations were carried out.

\bibliography{refs}

\section{End Matter}

\begin{figure}[t]
    \centering
    \includegraphics{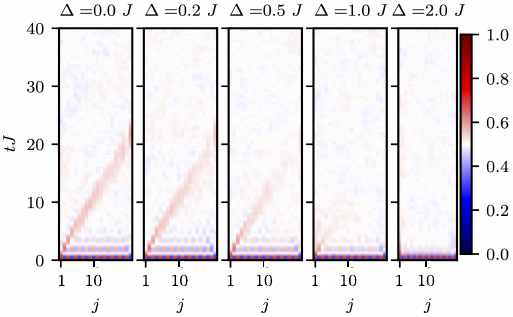}
    \caption{Dynamics of particle density with OBC for different values of the interaction $\Delta$ and for $\phi=-\theta=\pi/2$, corresponding to non-reciprocal hoppings $J_+\neq J_-$ in the single particle non-Hermitian Hamiltonian controlling the unconditional evolution. Other parameters: $L=20$ and $\Gamma=\kappa=0.1J$}
    \label{fig:directional_dynamics_interaction}
\end{figure}

\emph{Impact of Interactions on Directional Dynamics under OBC --} We discuss the impact of interactions on the directional dynamics observed under OBC along the $\phi=-\theta$ line, corresponding to non-reciprocal hoppings $J_+\neq J_-$ in the non-interacting model (see Fig.~\ref{fig:non_interacting}(d)). To this end, we plot in Fig.~\ref{fig:directional_dynamics_interaction} the dynamics of particle density for different values of the interaction $\Delta$, starting from a CDW state. We see that the CDW pattern quickly melts after a few coherent revivals at short times. Furthermore, the directional dynamics observed in the non-interacting case, with a bump in the density profile traveling from the left edge of the system, remains well-defined for moderate values of the interaction ($\Delta \lesssim J$). The effect of $\Delta$ is to increase the damping of this mode, which, already for $\Delta=0.5 J$, does not travel far enough to reach the boundary and dissipates in the bulk of the chain.  For $\Delta>J$, on the other hand, the directionality of the dynamics is barely visible, and the initial CDW state rapidly approaches a uniform steady state with no net particle current. As we analytically show in~\cite{supplementary_mat}, the dynamics heats up the state to infinite temperature independently of $\Delta$.

\emph{Entanglement Entropy of Quantum Trajectories --} Here we discuss the entanglement structure of our model of non-reciprocal driven-dissipative fermions. The density matrix $\rho(t)$ solution of the Lindblad master equation defined in the main text describes, in general, a mixed state. In order to characterize entanglement and quantum correlations, we therefore use an unraveling of the system density matrix $\rho(t)$ into many-body quantum trajectories~\cite{dalibard1992wavefunction,plenio1998quantum,daley2014quantum,fazio2025manybodyopenquantumsystems}. These describe the non-unitary, but purity-preserving, evolution of the system conditioned to a series of measurement outcomes associated to the set of jump operators $L_{j\mu}$. Given a quantum trajectory $\vert\psi(\xi_t,t)\rangle$ evolving under the quantum jumps dynamics, we introduce the conditional density matrix $\rho(\xi_t,t)=\vert\psi(\xi_t,t)\rangle\langle \psi(\xi_t,t)\vert$. Upon averaging this quantity over the measurement noise $\xi_t$, one recovers the mixed density matrix solution of the Lindblad master equation, i.e. $\rho(t)=\overline{\vert \psi(\xi_t,t)\rangle\langle  \psi(\xi_t,t)\vert}$. 
\begin{figure}
\centering
\includegraphics{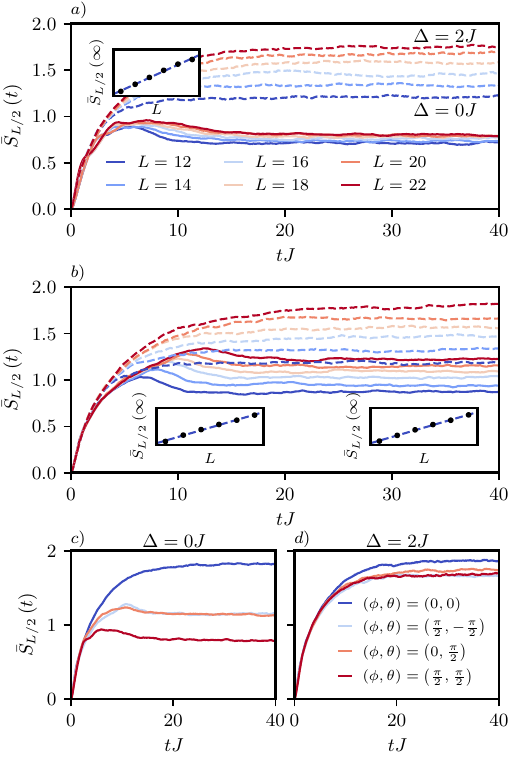}\caption{$a)$ and $b)$ - Time evolution of the average entanglement entropy for different system sizes and interaction strength with $\phi=\theta=\pi/2$ $[a)]$ and $\phi=-\theta=\pi/2$ $[b)]$. The full lines corresponds to the non-interacting case, while the dashed line corresponds to $\Delta=2.0J$. $c)$ and $d)$ - Time evolution of the entanglement entropy for different values of the angles for $L=20$. Panel $c)$ corresponds to the non-interacting case, while in $d)$ $\Delta=2.0J$. The label is shared between panels $a)$ and $b)$ and panels $c)$ and $d)$. Other parameters: $\Gamma=\kappa=0.1J$.}
\label{fig:entanglement}
\end{figure}
As long as one is interested in standard average quantities $\langle O\rangle=\mbox{Tr}\left(\rho(t)O\right)$, the conditional and average (also called unconditional) state contain the same information. However, this is not the case for quantities that involve non-linear functions of the state, for instance, probing higher moments of the conditional density matrix $\rho(\xi_t,t)$. A relevant example is provided by the von Neumann entanglement entropy, defined as~\cite{amico2008entanglementinmanybody}
\begin{equation}
S_A(t,\xi_t )  = -\mathrm{tr}_A\left[ \rho_\mathbf{\xi}^A(t)\ln \rho_\mathbf{\xi}^A(t)\right],
\end{equation}
where we have introduced a bipartition $A\cup B$ of the system and defined the reduced density matrix $\rho^A(\xi_t,t) = \mathrm{tr}_B |\psi(\xi_t,t)\rangle\langle\psi(\xi_t,t)|$. In the following, we are interested in the average entanglement entropy, given by $\bar{S}_A (t) = \int \mathcal{D}\mathbf{\xi}\; P(\mathbf{\xi}) S_A(\xi_t,t)$ where the average is taken over the Poisson measurement noise $\xi_t$. Clearly, the average entanglement entropy does not generally coincide with the entanglement entropy of the average state. The former is known to undergo a non-trivial evolution under continuous monitoring, leading to measurement-induced phase transitions.

In the present context, we focus on the effect of interaction and non-reciprocity on the dynamics of the entanglement entropy. In Fig.~\ref{fig:entanglement}(a) we plot the time-evolution of the average entanglement entropy for a bipartition of size $L/2$, different system sizes at fixed angles $\phi=\theta=\pi/2$ and for both $\Delta=0$ (non-interacting fermions) and $\Delta=2J$.  Interestingly, we note that while in the non-interacting case the entanglement entropy has a very weak dependence on system size and saturates to a constant area-law value, for $\Delta\neq0$ the entanglement dynamics clearly depends on the size of the subsystem. At long-times, the entanglement saturates to a steady-state value that depends linearly on $L$ (see inset), i.e. the scaling is compatible with a volume-law. For interacting monitored quantum many-body systems, volume-law scaling of entanglement entropy is expected, at least for weak monitoring~\cite{xing2023interactions,poboiko2025measurement}. However, this is rather surprising in the present context of non-reciprocal fermions, given the fact that from the point of view of the steady state the charge is localized at the boundary as in the skin-effect~\cite{kawabata2023entanglement}, as shown in Fig.~\ref{fig:dynamics_interaction}. We contrast this behavior with the entanglement entropy scaling along the critical lines $\phi=-\theta=\pi/2$, shown in Fig.~\ref{fig:entanglement}(b) which already in the non-interacting case displays a non-trivial system size dependence, likely reflecting the criticality of the modes along this line. Adding many-body interactions increases the entanglement entropy, which displays now a robust volume law scaling similar to what is shown in panel a). Finally, we explore the dependence of the entanglement entropy dynamics from the non-reciprocity, by tuning the angles $\phi,\theta$ for $\Delta=0$ (Fig.~\ref{fig:entanglement}(c)  and $\Delta=2J$ (Fig.~\ref{fig:entanglement}(d). By comparing the two panels, we first note that in the fully reciprocal case, corresponding to $\phi=\theta=0$, the entanglement dynamics depends only weakly on the interaction $\Delta$. Non-reciprocity, on the other hand, suppresses the entanglement growth in the non-interacting case, an effect which is significantly less pronounced in the interacting non-reciprocal case, whose entanglement content appears to be more robust.

\nocite{Haga2021Liouvillian,kamenev2023field,thompson2023field,Landi2024_prx_quantum}
\end{document}


\date{\today}
	
	\title{Supplemental Material to `Dissipative phase transition of interacting non-reciprocal fermions'}
	
	\author{Rafael D. Soares}
	\affiliation{Max Planck Institute for the Physics of Complex Systems, N\"{o}thnitzer Stra{\ss}e 38, 01187 Dresden, Germany}
	\affiliation{JEIP, UAR 3573 CNRS, Coll\`ege de France, PSL Research University, 11 Place Marcelin Berthelot, 75321 Paris Cedex 05, France}
	\author{Matteo Brunelli}
	\affiliation{JEIP, UAR 3573 CNRS, Coll\`ege de France, PSL Research University, 11 Place Marcelin Berthelot, 75321 Paris Cedex 05, France}
	\author{Marco Schir\`o}
	\affiliation{JEIP, UAR 3573 CNRS, Coll\`ege de France, PSL Research University, 11 Place Marcelin Berthelot, 75321 Paris Cedex 05, France}

\maketitle

\onecolumngrid

\renewcommand{\thefigure}{S\arabic{figure}}

\renewcommand*{\citenumfont}[1]{S#1}
\renewcommand*{\bibnumfmt}[1]{[S#1]}

\newcounter{ssection}
\stepcounter{ssection}

\setcounter{table}{0}
\setcounter{page}{1}
\setcounter{figure}{0}
\setcounter{equation}{0}

\makeatletter
\renewcommand{\theequation}{S\arabic{equation}}
\tableofcontents

\section{Symmetries of the Lindblad Master Equation}

We briefly discuss the symmetries associated with the Lindblad master equation and the jump operators defined in the main text. In addition to a standard weak $U(1)$ symmetry,
We briefly discuss the symmetries associated with the Lindblad master equation and the jump operators defined in the main text. In addition to a standard weak $U(1)$ symmetry, where the fermionic operators can be rotated by an arbitrary global phase $c_j\rightarrow e^{i\alpha}c_j$ without changing the master equation, the system also features a local gauge symmetry associated with the phases $\phi,\theta$. In particular under PBC and local transformation
\begin{equation}
c_j\rightarrow e^{i r_j\theta }c_{j}, \quad c^{\dagger}_j\rightarrow e^{-i r_j\theta }c^{\dagger}_{j}
\end{equation}
where $r_j$ is the position of the site $j$, one can eliminate the phase $\theta$ from the gain term, whose jump operators transform as
\begin{align}
L_{j,g}\rightarrow \sqrt{\Gamma}e^{-i r_j\theta}\left(c^{\dagger}_j+c^{\dagger}_{j+1}\right),
\end{align}
giving rise to a gain-dissipator which is independent of $\theta$. At the same time, the loss term acquires a $\theta$ dependence, 
\begin{align}
L_{j,\ell}\rightarrow \sqrt{\kappa}e^{ir_j\theta}\left(c_j+e^{i(\phi+\theta)}c_{j+1}\right),
\end{align}
and similarly, the hopping term acquires a Peierls-like phase 
\begin{equation}
-\dfrac{J}{2}\left( c^{\dagger}_jc_{j+1}+\text{h.c.}\right)\rightarrow
-\dfrac{J}{2}\left( e^{i\theta}c^{\dagger}_jc_{j+1}+\text{h.c.}\right).
\end{equation}
In other words, our model with non-reciprocal gain and losses is equivalent to a model with momentum-dependent (but reciprocal) gain, non-reciprocal losses with a phase $\theta+\phi$ and non-reciprocal hopping due to a Peierls phase $e^{i\theta}$. However, we note that, while the phase $\theta$ can be eliminated from the Lindbladian, it still affects physical observables which need to be transformed as well.
Finally, we stress that our model in the absence of interaction $\Delta=0$ has particle-hole symmetry for $\Gamma=\kappa$ and $\phi=\theta$ since in this case the master equation is invariant under $c_j\leftrightarrow c^{\dagger}_j$. A finite $\Delta$ gives rise to a shift in the Hamiltonian $\delta H=-2\Delta \sum_i n_i$.

\section{Non-Interacting Case: Equations of Motion and Continuity Equation}

In this section, we discuss in more detail the non-interacting limit ($\Delta=0$). We provide additional details that support the results discussed in the main text for both open-boundary and periodic boundary conditions. First, we write the equation of motion for the point correlation function,
\begin{equation}
\begin{aligned}i\partial_{t}\left\langle c_{n}^{\dagger}c_{m}\right\rangle  & =-\dfrac{1}{2}\left[J+i\kappa e^{i\phi}+i\Gamma e^{-i\theta}\right]\left\langle c_{n}^{\dagger}c_{m+1}\right\rangle -\dfrac{1}{2}\left[J+i\kappa e^{-i\phi}+i\Gamma e^{i\theta}\right]\left\langle c_{n}^{\dagger}c_{m-1}\right\rangle +\\
 & +\dfrac{1}{2}\left[J-i\kappa e^{-i\phi}-i\Gamma e^{i\theta}\right]\left\langle c_{n+1}^{\dagger}c_{m}\right\rangle +\dfrac{1}{2}\left[J-i\kappa e^{i\phi}-i\Gamma e^{-i\theta}\right]\left\langle c_{n-1}^{\dagger}c_{m}\right\rangle+\\
 & -2i\kappa\left\langle c_{n}^{\dagger}c_{m}\right\rangle +2i\Gamma\left(\delta_{n,m}-\left\langle c_{n}^{\dagger}c_{m}\right\rangle \right)+\\
 & +i\dfrac{\Gamma}{2} \left(e^{-i\theta}\left[\delta_{n,m+1}+\delta_{n-1,m}\right]+e^{i\theta}\left[\delta_{n,m-1}+\delta_{n+1,m}\right] \right).
\end{aligned}
\label{eq:equation_of_motion_no_interacting_case}
\end{equation}
As already discussed, both gain and loss dissipation couple to the hopping elements of the unitary evolution, leading to the appearance of two distinct effective hopping rates: $J_\pm=J + i\kappa e^{\pm i \phi} + i\Gamma e^{\mp i\theta}$.
It is interesting to see how these effective couplings affect the dynamics of the particle density at a site $n$,
\begin{equation}
\partial_{t}\left\langle c_{n}^{\dagger}c_{n}\right\rangle + v_+ \langle \mathcal{I}_{n}\rangle - v_-\langle\mathcal{I}_{n-1} \rangle+ \dfrac{\langle h_{n}\rangle + \langle h_{n-1} \rangle}{2J}\left[\kappa\cos\left(\phi\right)+\Gamma\cos\left(\theta\right)\right]=\langle \mathcal{T}_{n} \rangle,
\label{eq:continuity_equation}
\end{equation}
where $ v_\pm = 1\pm\left[\Gamma/J\sin\left(\theta\right)-\kappa/J\sin\left(\phi\right)\right]$, $\langle \mathcal{T}_n \rangle =2\Gamma-2\left(\Gamma+\kappa \right)\langle c^\dagger_n c_n \rangle$ which is the global rate of particle injection/losses from the bath, $\langle \mathcal{I}_n\rangle$ the expectation value of the local particle current operator at a site $n$ and $\langle h_n\rangle$ the expectation value of the local kinetic energy density at a site $n$, namely,
\begin{equation}
\langle \mathcal{I}_{n} \rangle=\dfrac{iJ}{2}\left(\left\langle c_{n+1}^{\dagger}c_{n}\right\rangle -\left\langle c_{n}^{\dagger}c_{n+1}\right\rangle \right), \quad\langle  h_{n} \rangle  = \langle c_{n+1}^{\dagger}c_{n}\rangle  +  \langle c_{n}^{\dagger}c_{n+1}\rangle.
\end{equation}
Eq.~\eqref{eq:continuity_equation} would describe a local continuity equation associated with global conservation of the total particle number if the system were governed solely by Hamiltonian dynamics. However, the right-hand side of the equation reveals that particles can be injected into or removed from the system due to coupling with the bath. In particular, the imaginary components of the effective hoppings $J_\pm$, renormalize and induce direction-dependent effective velocities. This is similar to what was observed in certain regimes for the no-click dynamics of the Hatano-Nelson model~\cite{Soares2024Faber}. In addition, their real part introduces a non-trivial contribution that depends on the local kinetic energy. 

As we show next, an imbalance in $v_\pm$ serves as a reliable indicator of directionality in the propagation of the local particle density. This is further evident from the local particle current in the PBC case, where, despite the absence of directional propagation - since the system ultimately reaches homogeneity - directionality can still be inferred from the presence of a net particle current.

\subsection{Results for Periodic Boundary Conditions}
\begin{figure}[t]
    \centering
    \includegraphics{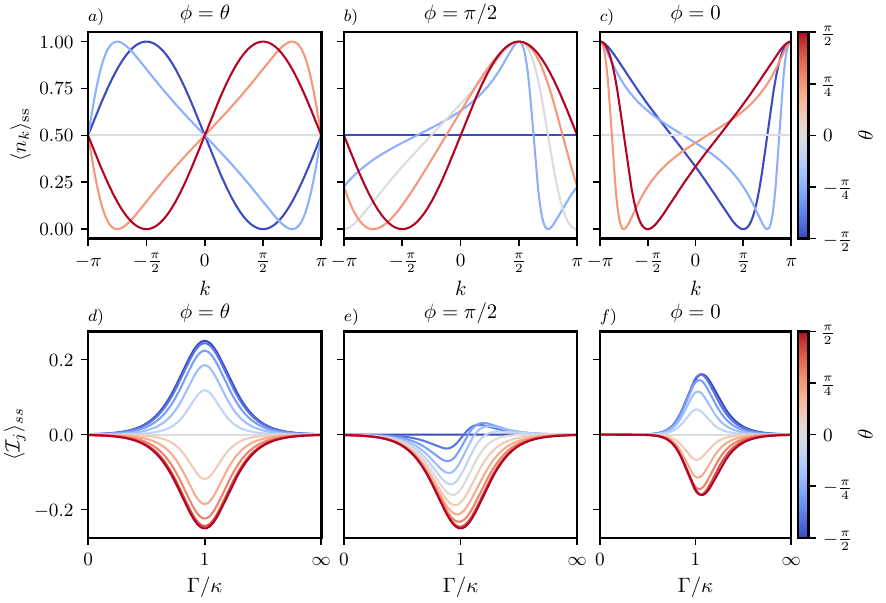}
    \caption{Upper row - Steady-state distribution for different values of the angle $\theta$ with $\phi$ fixed with $\Gamma=\kappa$. Lower row - Steady-state particle current as a function of the ratio $\Gamma/\kappa$ for different values of the angle $\theta$ with $\phi$ fixed. The $\Gamma/\kappa$ axis is in log-scale.}
    \label{fig:steady_state_nk_current_pbc_vs_ratio}
\end{figure}
Here we provide further analytical results for the PBC case in the absence of interactions. In this case, the equation of motion for the correlation matrix Eq.~\eqref{eq:equation_of_motion_no_interacting_case}, when written in momentum space $C_{kq}=\left\langle c_{k}^{\dagger}c_{q}\right\rangle$, simplify to
\begin{equation}
\partial_{t}C_{kq} =-\left[i\left(\varepsilon_k -\varepsilon_q \right)+\dfrac{\lambda_k+\lambda_q}{2} \right]C_{kq} +\delta_{k,q}\: \Gamma_k,
\end{equation}
with $\lambda_k=\Gamma_k + \kappa_k$ as defined in the main text, and $\varepsilon_k=-J\cos\left(k \right)$. The solution corresponds, for the off-diagonal matrix elements, to an exponential decay in time, 
\begin{equation}
C_{kq}(t)=C_{kq}(0)e^{-i\left(\varepsilon_k -\varepsilon_q \right)t}e^{-(\lambda_k+\lambda_q)t/2},
\end{equation}
while, for the diagonal elements of the correlation matrix (that is the distribution function $n_k(t)=C_{kk}(t)$), we obtain Eq.~(6) of the main text.
First, we discuss the steady-state solution of these equations of motion. In the steady-state, the correlation matrix is diagonal in momentum; as such, the steady-state particle density and current are spatially uniform and given by
\begin{equation}
\left\langle n_j\right\rangle_{\text{ss}} =\int_{-\pi}^{+\pi}\dfrac{dk}{2\pi} \langle n_{k} \rangle_{\text{ss}},\qquad
\left\langle \mathcal{I}_{j}\right\rangle_{\text{ss}}
=-J\int_{-\pi}^{+\pi}\dfrac{dk}{2\pi}\sin\left(k\right) \langle n_{k} \rangle_{\text{ss}},
\end{equation}

where the steady-state distribution is given in Eq.~(8)
of the main text, that we rewrite here as a function of the angles $\theta,\phi$ and the ratio between the gain and loss rate $\kappa/\Gamma$
\begin{equation}\label{eqn:distribution}
\langle n_{k} \rangle_{\text{ss}} =\dfrac{\left(1+\cos\left(k-\theta\right)\right)}{\left(1+\cos\left(k-\theta\right)\right)+\dfrac{\kappa}{\Gamma}\left(1+\cos\left(\phi+k\right)\right)}.
\end{equation}

In Fig.~\ref{fig:steady_state_nk_current_pbc_vs_ratio}, we plot the distribution function $\langle n_{k}\rangle_{\text{ss}}$ for different values of the angles (top row) and the corresponding particle current as a function of $\Gamma/\kappa$ and the two angles (bottom row). As mentioned in the main text, the steady state does not support a particle current when the momentum distribution $\langle n_k \rangle_{\rm ss}$ is flat in $k$,
which occurs along the line $\phi=-\theta$. In this case, as seen from Eq.~\eqref{eqn:distribution}, the two dissipative processes interfere destructively and cancel out, leaving a distribution function $\langle n_{k}\rangle_{\text{ss}}=\Gamma/(\Gamma+\kappa)$, which is identical to the case of local onsite gain ($L_{j,g}=\sqrt{\Gamma}c^\dagger_{j}$) and losses ($L_{j,\ell}=\sqrt{\kappa}c_{j}$). In contrast, the particle current is maximized when $\langle n_{k}\rangle_{\text{ss}} = (1 + \sin(k))/2$, which occurs for $\theta = \phi = \pm\pi/2$ and $\Gamma = \kappa$, as clearly seen in the lower row of Fig.~\ref{fig:steady_state_nk_current_pbc_vs_ratio} and in Fig.~(2).(b) 
of the main text. In this case, the modes with the highest group velocity are the most populated, resulting in a maximal current. As the angles are tuned away from this line, the distribution function loses the (anti)symmetry with respect to inversion, thus reducing the current. Moreover, as shown in the figure, the particle current vanishes in both limits $\Gamma/\kappa \to 0$ and $\Gamma/\kappa \to \infty$, corresponding to a fully empty and fully filled chain, respectively, and reaches its maximum at $\Gamma = \kappa$. The sign of the current is controlled by both angles but also by the gain/loss ratio. This is seen in Fig.~\ref{fig:steady_state_nk_current_pbc_vs_ratio}.(e), where for some angles $\theta$ and a fixed $\phi=\pi/2$, the current can be reversed by increasing the gain/loss ratio. When $\theta=\phi$, we can derive an analytically expression for the particle current, also assuming that $\Gamma=\kappa$,
\begin{equation}
\langle\mathcal{I}_{j}\rangle_{\text{ss}} = J\dfrac{\sin(\theta) }{2\cos^2(\theta)}\left(|\sin(\theta)|-1 \right),
\end{equation}
in this case, the direction of the particle current is given by the sign of the $\sin\left(\theta\right)$.

\begin{figure}[t]
    \centering
    \includegraphics{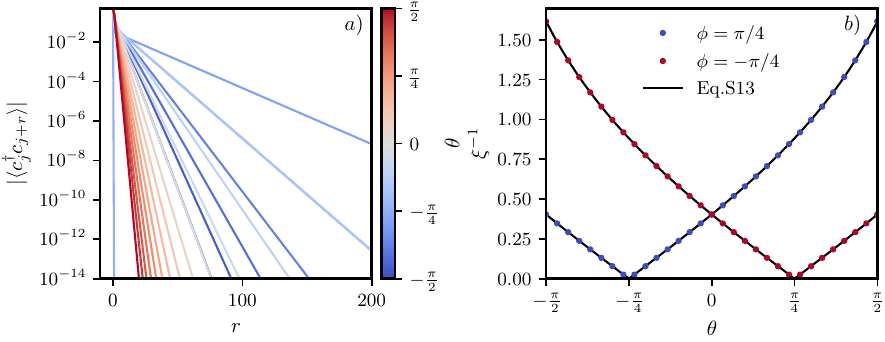}
    \caption{a) - Absolute value of the two-point correlation function as a function of distance $r$ for $\phi=\pi/4$. The correlation length is minimal and equal to zero for $\theta=-\phi$. b) - Inverse correlation length as a function as function of $\theta$ for two distinct values of $\phi$. Other parameters: $\Gamma=\kappa=0.1J$}
    \label{fig:steady_state_correlation_length}
\end{figure}

It is also interesting to explore other correlation functions. We note that the steady state is never critical: the steady-state two-point correlation function always decays exponentially, $\langle c^\dagger_{j} c_{j+r} \rangle \propto e^{-r/\xi}$, with a correlation length $\xi$ that remains finite throughout the phase diagram for all system parameters. This correlation length depends sensitively on the angles and exhibits a non-monotonic dependence as one angle is varied relative to the other, as shown in Fig.~\ref{fig:steady_state_correlation_length}.(a). This can be analytically determined through the poles of the distribution $\langle n_k \rangle_{\text{ss}}$, which indicate that
\begin{equation}
    \xi^{-1}= \dfrac{1}{2}\log\left(\dfrac{\Gamma+\kappa+2\sqrt{\Gamma\kappa}\left|\sin\left(\dfrac{\theta+\phi}{2}\right)\right|}{\Gamma+\kappa-2\sqrt{\Gamma\kappa}\left|\sin\left(\dfrac{\theta+\phi}{2}\right)\right|} \right).
\end{equation}
We note that this expression is not valid for $\theta=-\phi$. As one approaches the line $\theta=-\phi$ the dissipative gap $\delta$ decreases toward zero, leading to an increasing correlation length. However, right at the critical line, $\delta=0$, also the gain vanishes in the same way, giving rise to a completely flat momentum distribution in $k$ space, i.e. $\langle c^\dagger_{j} c_{j+r} \rangle=\delta_{r,0}$.

For what concerns the transient dynamics, we first consider an initially empty lattice such that the dynamics of particle density and current does not depend on initial off diagonal correlations in momentum space. As such, the particle density and coherent current at a site $j$ can be formally obtained by using,
\begin{align}
\left\langle n_j(t)\right\rangle  =\int_{-\pi}^{+\pi}\dfrac{dk}{2\pi} \langle n_{k}(t) \rangle,\,\qquad
\left\langle \mathcal{I}_{j}(t)\right\rangle
=-J\int_{-\pi}^{+\pi}\dfrac{dk}{2\pi}\sin\left(k\right) \langle n_{k}(t) \rangle,
\end{align}
where the time dependent distribution function is given in Eq.~(6) of the main text.  
We start by considering the line $\phi=-\theta$, for which we have $\lambda_k = 2\left(\Gamma+\kappa\right)\left[1+\cos\left(k+\phi\right)\right]$. In this case we can rewrite the particle density as
\begin{equation}
\begin{aligned}
\left\langle n_j(t)\right\rangle  
 & =\dfrac{\Gamma}{\Gamma+\kappa}\left(1-e^{-2\left(\Gamma+\kappa\right)t}\int_{-\pi}^{+\pi}\dfrac{dk}{2\pi}e^{-2\left(\Gamma+\kappa\right)\cos\left(k+\phi\right)t}\right)\\
 & =\dfrac{\Gamma}{\Gamma+\kappa}\left(1-e^{-2\left(\Gamma+\kappa\right)t}\sum_{n=-\infty}^{+\infty}\int_{-\pi}^{+\pi}\dfrac{dk}{2\pi}\left(-ie^{i(k+\phi)}\right)^{n}J_{n}\left(2i\left(\Gamma+\kappa\right)t\right)\right)\\
 & =\dfrac{\Gamma}{\Gamma+\kappa}\left(1-e^{-2\left(\Gamma+\kappa\right)t}I_{0}\left(2\left(\Gamma+\kappa\right)t\right)\right),
 \end{aligned}
 \label{eq:density_pbc_vaccum}
\end{equation}
where $I_{0}(\cdot)$ is the zero order Bessel Function of the second kind. Similarly, for the current, we get
\begin{equation}
\langle\mathcal{I}_{j}\left(t\right)\rangle  =-\dfrac{\Gamma}{\Gamma+\kappa} I_{1}\left(2(\Gamma + \kappa) t\right)e^{-2(\Gamma + \kappa) t}\sin\left(\phi\right),
\end{equation}
with $I_{1}(\cdot)$ the first order Bessel Function of the second kind.
As mentioned in the main text, we see that there is a transient current as long as $\phi\neq n\pi$, even though in the steady-state the current vanishes. We also see that the direction is controlled by the angle $\phi$.

In the limit $t\rightarrow + \infty$, we can expand the Bessel functions and obtain the asymptotics of particle density and current
\begin{equation}
\begin{aligned}
\left\langle n_{j} (t\rightarrow\infty)\right\rangle &= \frac{\Gamma}{\Gamma+\kappa} \left(1-\frac{1}{2\sqrt{\pi}\sqrt{\Gamma+\kappa}}\sqrt{\frac{1}{t}}\right) + O\left(t^{-3/2}\right),\\
\langle\mathcal{I}_{j} (t\rightarrow\infty)\rangle &= \dfrac{\Gamma}{2\sqrt{\pi}(\Gamma+\kappa)^{3/2}} \sin\left(\phi\right) \sqrt{\frac{1}{t}}+O\left(t^{-3/2}\right).
\end{aligned}
\end{equation}
In both cases, we obtain a power-law decay towards the steady-state value (which is zero for the current) of the form $1/t^{\alpha}$ with an exponent $\alpha=1/2$. This decay is ultimately due to the fact that the spectrum of decay rates $\lambda_k$ vanishes quadratically around $k_*$. On the other hand, away from the line $\theta=-\phi$, the spectrum $\lambda_k$ is gapped, $\lambda_k=\delta+\alpha(k-k_*)^2$ and the dynamics of both particle density and current acquire an additional exponential decay with a scale controlling the distance from the critical line $\phi=-\theta$.

\subsection{Dynamics for an initial charge density wave state}

In this subsection, we discuss the time evolution of the particle and the current density for an initial state in a charge-density wave (CDW), as done in the main text. When considering periodic boundary conditions, in the thermodynamic limit, the initial state correlation matrix in momentum space can be written as
\begin{equation}
    \langle c^\dagger_k c_q \rangle = \dfrac{1}{2}\delta_{k,q} + \dfrac{1}{2}\delta_{k+\pi,q}.
\end{equation}

As such, the particle density and particle current, for all angles, are given by,
\begin{equation}
\begin{aligned}
    \langle n_j \left(t \right) \rangle &=\langle n_j \rangle_{\rm ss} + \dfrac{1}{2}\int_{-\pi}^{\pi} \dfrac{dk}{2\pi} \left[ \left(1-\dfrac{2\Gamma_k}{\Gamma_k+\kappa_k}\right)e^{-\lambda_{k}t} + (-1)^j e^{-2\left(\Gamma+\kappa \right)t - 2 i \varepsilon_k t} \right],\\
    \langle \mathcal{I}_j \left(t \right) \rangle &=\langle \mathcal{I}_j \rangle_{\rm ss} - \dfrac{J}{2}\int_{-\pi}^{\pi} \dfrac{dk}{2\pi} \left[ \left(1-\dfrac{2\Gamma_k}{\Gamma_k+\kappa_k}\right)e^{-\lambda_{k}t}\sin(k) + (-1)^j e^{-2\left(\Gamma+\kappa \right)t}\sin \left(k-2\varepsilon_kt \right)\right] .
\end{aligned}
\end{equation}
These integrals can be analytically solved in the same way as before (Eq.~\eqref{eq:density_pbc_vaccum}) by using the Jacobi–Anger expansion.
\begin{figure}
    \centering
    \includegraphics{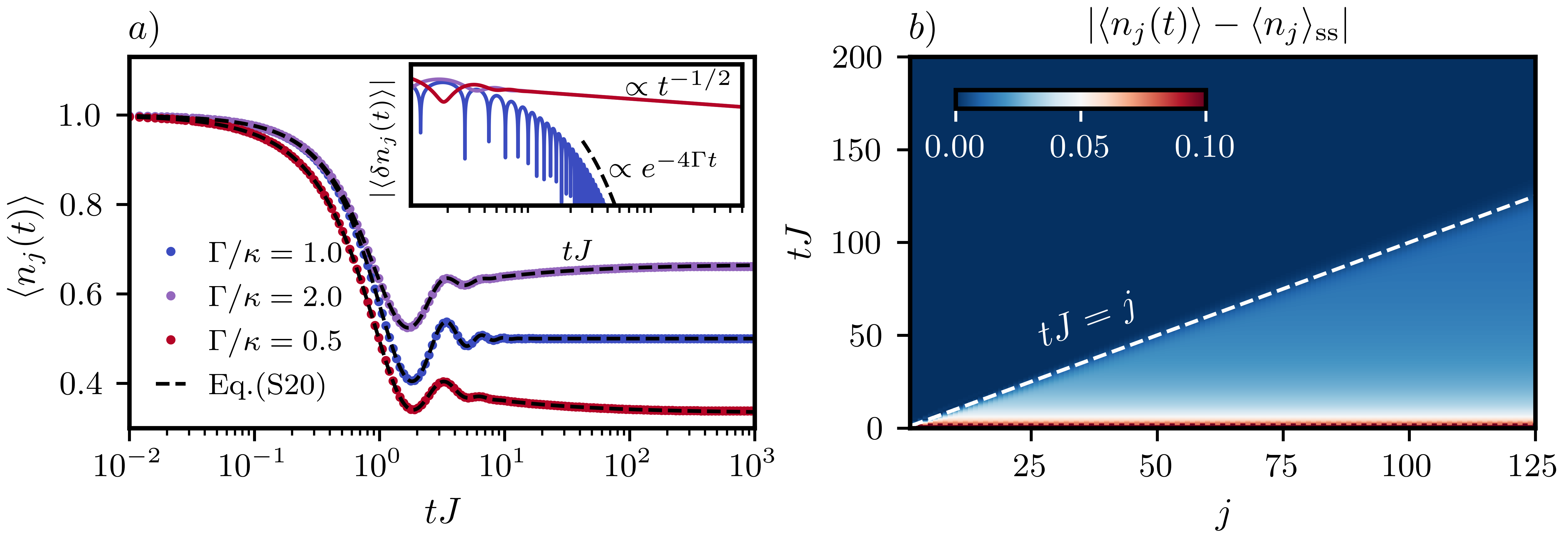}
    \caption{Time evolution of the local particle density from a CDW initial state. $a)$ Comparison of the analytical expression and numerical results for various ratios of $\Gamma/\kappa$ under PBC. The inset illustrates the approach to the steady state via $ \langle \delta n_j (t) \rangle =\langle n_j(t) \rangle - \langle n_{j} \rangle_{\rm ss}$. $b)$ Time evolution of the local particle density for $\Gamma = 0.5J$ and $\kappa = 0.1J$ in a system with OBC. Other simulation parameters: $\phi = -\theta = \pi/2$.}
    \label{fig:dynamics_cdw_2}
\end{figure}
First, we consider the regime where $\theta=-\phi$. Along this specific line of angles, the dynamics starting from an initial CDW for the local particle density and particle current are given by
\begin{equation}
\begin{aligned}
\langle n_j(t)\rangle  &=\dfrac{\Gamma}{\Gamma+\kappa}+\left[\dfrac{1}{2}-\dfrac{\Gamma}{\Gamma+\kappa}\right]e^{-2(\Gamma+\kappa)t}I_{0}\left(2\left[\Gamma+\kappa\right]t\right)+\dfrac{\left(-1\right)^{j}}{2}e^{-2\left(\Gamma+\kappa\right)t}J_{0}\left(2Jt\right), \\
\langle \mathcal{I}_j(t) \rangle &= -\left[\dfrac{1}{2}-\dfrac{\Gamma}{\Gamma+\kappa}\right] I_{1}\left(2(\Gamma + \kappa) t\right)e^{-2(\Gamma + \kappa) t}\sin\left(\phi\right)-\dfrac{(-1)^j}{2}e^{-2\left(\Gamma+\kappa \right)t}J_{1}\left(2Jt \right),
\end{aligned}
\end{equation}
where $I_\alpha(\cdot)$ and $J_\alpha(\cdot)$ are the $\alpha$-order modified Bessel function of the first kind and the $\alpha$-order Bessel function of the first kind, respectively. For all values of $\Gamma$ and $\kappa$, the initial charge density wave order vanishes exponentially fast, as seen by the second term in the equation. However, the way in which the particle density reaches the steady state can be altered by tuning the ratio $\Gamma/\kappa$. Specifically, in the long-time limit, both the particle density and current approach the steady state following a power law given by $t^{-1/2}$, as long as $\Gamma\neq\kappa$,
\begin{equation}
\begin{aligned}
\langle n_j(t\rightarrow\infty)\rangle - \langle n_j \rangle_{\rm ss} &=\frac{\kappa-\Gamma}{4\sqrt{\pi}\left(\Gamma+\kappa\right)^{3/2}}\sqrt{\frac{1}{t}}+O\left(t^{-3/2}\right),\\
\langle \mathcal{I}_j(t\rightarrow\infty)\rangle - \langle \mathcal{I}_j \rangle_{\rm ss} &=\frac{\Gamma-\kappa}{4\sqrt{\pi}\left(\Gamma+\kappa\right)^{3/2}} \sin\left(\phi \right)\sqrt{\frac{1}{t}}+O\left(t^{-3/2}\right).
\end{aligned}
\end{equation}
In contrast, when both $\Gamma=\kappa$, the relaxation is exponentially fast
\begin{equation}
\begin{aligned}
\langle n_j(t\rightarrow\infty)\rangle - \langle n_j \rangle_{\rm ss} &= \dfrac{\left(-1\right)^{j}}{2}e^{-2\left(\Gamma+\kappa\right)t}J_{0}\left(2Jt\right), \\
\langle \mathcal{I}_j(t\rightarrow\infty)\rangle - \langle \mathcal{I}_j \rangle_{\rm ss} &=-\dfrac{(-1)^j}{2}e^{-2\left(\Gamma+\kappa \right)t}J_{1}\left(2Jt \right),
\end{aligned}
\end{equation}
as we show in the inset of Figure~\ref{fig:dynamics_cdw_2}(a) for the particle density. For a generic initial state, one observes a power-law decay toward the steady state for all values of the ratio $\Gamma/\kappa$. The exponential behavior seen here is a unique consequence of our initial condition, since the diagonal elements of the correlation matrix are the same in the initial state and in the steady-state $\langle c^\dagger_k c_k(0) \rangle=\langle c^\dagger_k c_k\rangle_{\text{ss}}$. As a result, the average total number of particles remains conserved throughout the evolution, and the external bath only leads to the decay of the off-diagonal elements of the correlation matrix, $\langle c^\dagger_kc_q \rangle$.

We now discuss the dynamics with OBC. In this case, we see a clear front in the propagation of the particle density, which is consistent with recent results~\cite{Song2019ChiralDamping}. At short times, we see that both OBC and PBC show qualitatively similar dynamics; however, in the long-time limit, the system under open boundary conditions shows an extra exponential decay, as observed in the panel of Fig.~\ref{fig:dynamics_cdw_2}.

The model studied in this work belongs to the class of Lindbladians that display the Liouville skin-effect~\cite{Song2019ChiralDamping,Haga2021Liouvillian}, and so, as already stressed in previous works~\cite{Song2019ChiralDamping}, the spectrum and eigenmodes of the Lindblad superoperator under open and periodic conditions are substantially different, leading, as a consequence, to quite distinct dynamics.

\section{Keldysh Approach}
\label{sec:keldysh_section}
Here we discuss the solution of the non-interacting problem $\Delta=0$ with  Keldysh techniques and the resulting Green's functions describing dynamical correlations above the steady-state. We start rewriting the Lindblad master equation in the main text as a path integral over fermionic coherent fields defined on the Keldysh contour $\psi_{ja},\bar{\psi}_{ja}$, where $j$ is the lattice site and $a=\pm$ labels the contour branch~\cite{kamenev2023field,sieberer2023universalitydrivenopenquantum,fazio2025manybodyopenquantumsystems,thompson2023field}. We have
\begin{align}
\mathcal{Z}=\int \mathcal{D}\psi \, \mathcal{D}\bar{\psi} \,e^{i\mathcal{S}[\psi,\bar{\psi}]},
\end{align}
where the Keldysh action, in real space, reads
\begin{align}
\mathcal{S}\left[\psi,\bar{\psi}\right]=
\int dt\sum_{ja} a\bar{\psi}_{ja}i\partial_t \psi_{ja}-\left(\sum_{a}a H_a[\psi,\bar{\psi}]+iD[\psi,\bar{\psi}]\right),
\end{align}
with $H$ the Hamiltonian, written in terms of the coherent fields, and $D[\cdot]$, the dissipator which is given by
\begin{align}
D[\psi,\bar{\psi}]=\sum_{j\mu} \bar{L}_{j\mu-}L_{j\mu+}-\frac{1}{2}
\left(\bar{L}_{j\mu+}L_{j\mu+}+\bar{L}_{j\mu-}L_{j\mu-}
\right),
\end{align}
where $L_{j\mu a},\bar{L}_{j\mu a}$ are the jump operators on site $j$ and $\mu=\ell,g$ describes loss/gain processes. The action in momentum-frequency space takes the form
\begin{align}
\mathcal{S}[\psi,\bar{\psi}]&=
\sum_{k}\int d\omega \; \bar{\Psi}_k \begin{pmatrix}
        \omega-\varepsilon_k+i(\kappa_k-\Gamma_k)/2  & i \Gamma_k  \\
        -i\kappa_k  & -\omega+\varepsilon_k+i(\kappa_k-\Gamma_k)/2 
    \end{pmatrix} \Psi_k,
\end{align}
\begin{figure}[t]
    \centering
    \includegraphics{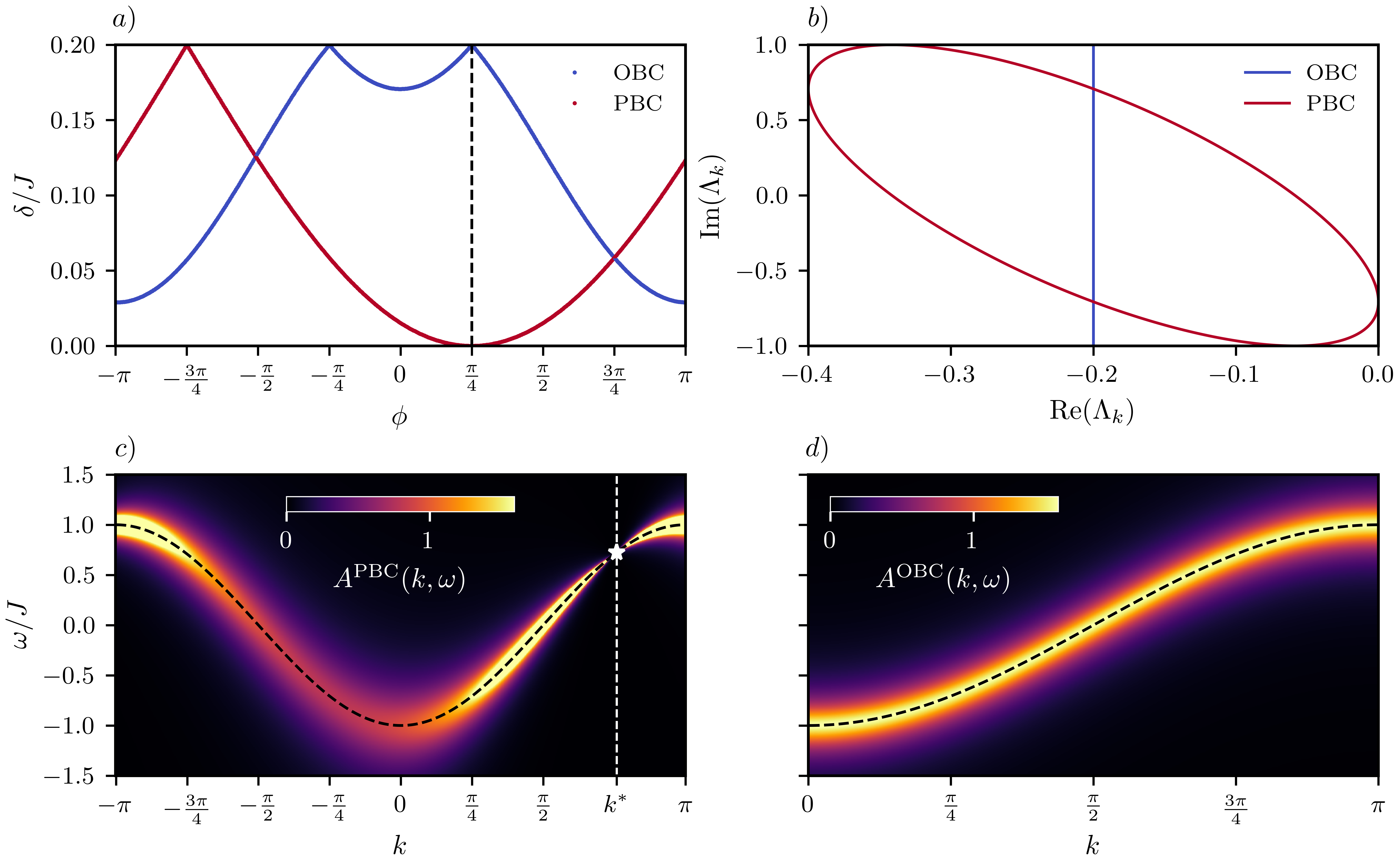}
    \caption{$a)$ Dissipative gap for OBC and PBC as a function of $\phi$ for $\theta=-\pi/4$. $b)$ Comparison of the single particle Linblad spectrum between OBC and PBC for $\phi=-\theta=\pi/4$/. $c)$ and $d)$ Spectral function for PBC (panel $c)$) and OBC (panel $d)$). In both panels, the dashed black line corresponds to the bare dispersion. In panel $c)$ the white star identifies the dark mode where the excitations above the steady-state are sharp coherent quasiparticles. Other parameters: $\Gamma = \kappa = 0.1J$.}
    \label{fig:dissipative_gap_obc_pbc}
\end{figure}
where we have defined the field $
\bar{\Psi}_k=\begin{pmatrix}
        \bar{\psi}_{k+} & \bar{\psi}_{k-}
    \end{pmatrix}$, and the momentum-dependent gain and loss rates are defined in the main text. 
From this action we can compute the Green's function in the $\pm$ basis, defined as
\begin{align}
\hat{G}_{k}(t,t')=-i\langle \Psi_k(t)\bar{\Psi}_k(t')\rangle=
\begin{pmatrix}
        G_k^{++} &  G_k^{+-}\\
        G_k^{-+} &  G_k^{--}\\
    \end{pmatrix} .
\end{align}
The Green functions in frequency/momentum space is given by the inverse of the Gaussian action and reads
\begin{equation}
\hat{G}(k,\omega)=
\begin{pmatrix}
        \omega-\varepsilon_k+i(\kappa_k-\Gamma_k)/2  & i \Gamma_k  \\
        -i\kappa_k  & -\omega+\varepsilon_k+i(\kappa_k-\Gamma_k)/2 
    \end{pmatrix}^{-1},
\end{equation}
It is convenient to introduce the retarded, advanced and Keldysh components, $G^{R,A,K}(k,\omega)$ which are directly related to the spectrum and the occupation of the excitations above the steady-state~\cite{kamenev2023field}. These are given by
\begin{equation}
    G^{R/A}(k,\omega)=\dfrac{1}{\omega-\varepsilon_k\pm i\lambda_k/2},\qquad G^K(k,\omega) =\dfrac{i\left(\Gamma_k-\kappa_k\right)}{(\omega-\varepsilon_k)^2+\lambda^2_k/4},
\end{equation}
From the retarded Green function, we can compute the spectral function $A(k,\omega)=-(1/\pi)\mbox{Im}G^R(k,\omega)$ of the fermions,
\begin{equation}
A(k,\omega)=\dfrac{1}{2\pi}\dfrac{\lambda_k}{(\omega-\varepsilon_k)^2+\lambda^2_k/4},
\end{equation}
which takes the form of a Lorentzian with a momentum-dependent broadening given by the spectrum of decay rates $\lambda_k$ defined in the main text. This broadening, which reflects the finite lifetime of excitations due to the coupling with the external baths, is minimal at $k_*(\theta,\phi)$ and equal to the dissipative gap, while it vanishes along the critical line $\theta=-\phi$. Here, the excitations above the steady-state are sharp coherent quasiparticles
\begin{equation}
    A(k^\ast,\omega)=\delta\left(\omega - \varepsilon_{k^\ast} \right).
\end{equation}
The dissipative transition in the dynamics discussed in the main text appears therefore as a transition in the spectral function of the fermions. This also translates into a transition in the decay of the local retarded Green's function in the time-domain, $G^R(x=0,t)=\int dk d\omega e^{-i\omega t}G^R(k,\omega)$ from exponential for $\theta\neq -\phi$ to power-law $G^R(x=0,t)\sim 1/\sqrt{t}$. We note that the decay is slower than the usual $~1/t$ decay of free fermions.

The Keldysh component of the Green function allows us to define an out of equilibrium distribution function $f(k,\omega)$ from an effective fluctuation-dissipation theorem~\cite{kamenev2023field}
\begin{equation}
f(k,\omega)=-\dfrac{G^K(k,\omega)}{2\pi i A(k,\omega)}=\dfrac{\kappa_k-\Gamma_k}{\lambda_k}.
\end{equation}
and consequently an effective temperature which, in general, depends on both $k$ and $\omega$ and it is defined by the relation
\begin{equation}
    \dfrac{\kappa_k-\Gamma_k}{\lambda_k}=\tanh\left(\dfrac{\omega}{2T_{\rm eff}} \right).
\end{equation}
However, we observe that along the critical line $\phi=-\theta$ with $\Gamma=\kappa$, $T_{\rm eff} \rightarrow +\infty$

Equally we can compute the retarded Green function under OBC by diagonalizing the effective non-Hermitian Hamiltonian associated to the unconditional evolution, which corresponds to an Hatano-Nelson type of Hamiltonian with the hopping given by $J_\pm$~\cite{kawabata2023entanglement,Soares2024Faber}. From this we obtain,
\begin{equation}
    G^R(k,\omega) = \dfrac{1}{\omega - \tilde{J} \cos(k) - i(\kappa+\Gamma)},
\end{equation}
with $k$ now quantized as $k=\pi\cdot n/(L+1),$ $n\in\{1,\cdots, L\}$ and
\begin{equation}
    \tilde{J} = \left[J^2-\Gamma ^2 - \kappa^2 + 2 \Gamma  \kappa  \cos (\theta +\phi ) + 2iJ\left(\kappa\cos(\phi) - \Gamma \cos(\theta) \right) \right]^{1/2}.
\end{equation}
In stark contrast to the PBC case, the dissipative gap remains finite for $\phi=-\theta$, as illustrated in Fig.~\ref{fig:dissipative_gap_obc_pbc} $(a)$. Furthermore, Fig.~\ref{fig:dissipative_gap_obc_pbc} $(b)$ reveals that the Lindblad spectrum is fundamentally altered by the boundary conditions. This sensitivity is also evident in the spectral function: while the excitation lifetime for $\phi=-\theta$ is momentum-dependent and diverges at $k^\star$, the lifetime under OBC becomes essentially independent of both momentum and frequency.

\section{Stability against Disorder and Dephasing}
\label{sec:disorder_dephasing}
\begin{figure}[t]
    \centering
    \includegraphics{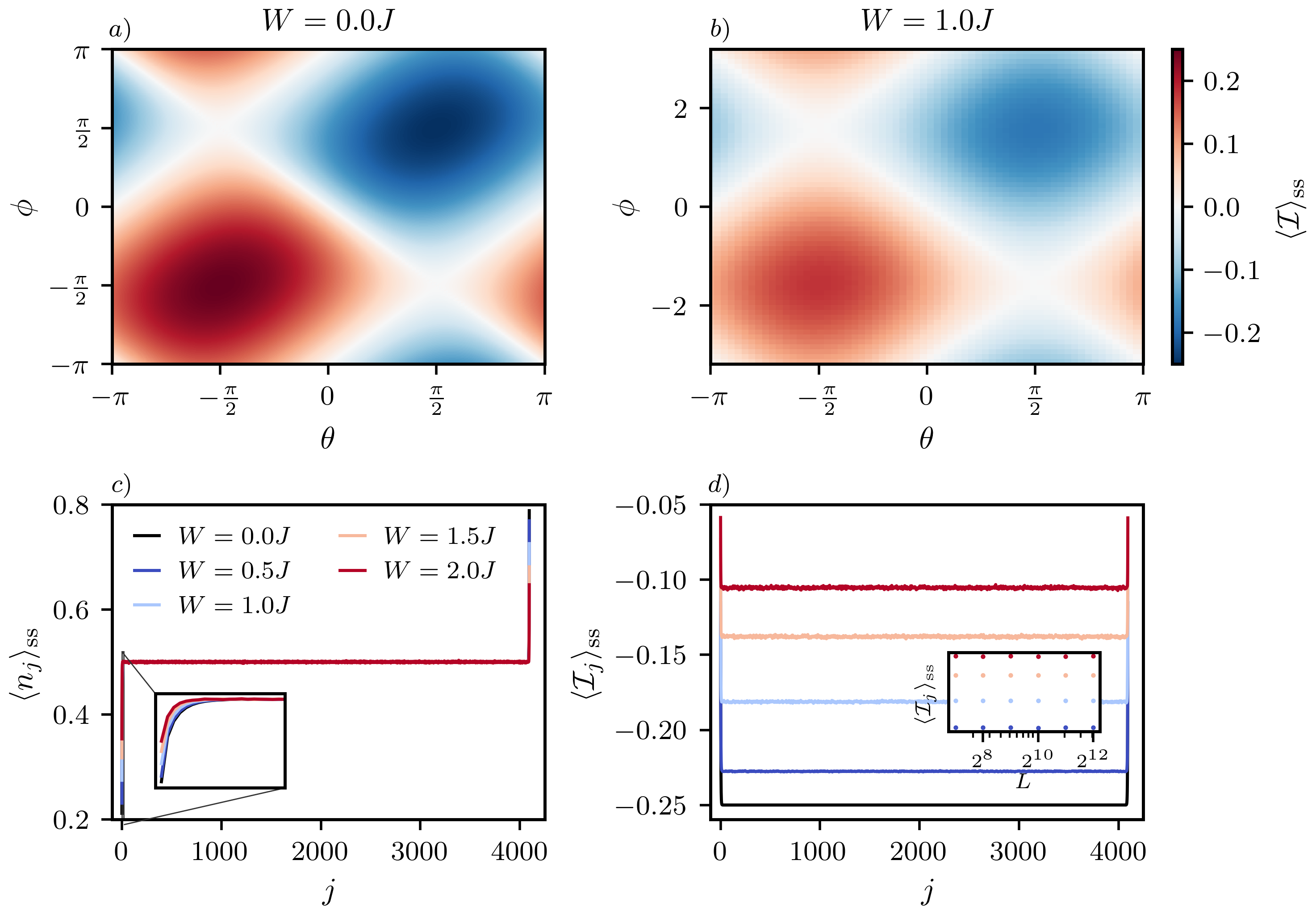}
    \caption{$a)$ Steady-state current as function of $\phi$ and $\theta$ for a clean system and $b)$ for disorder strength $W=1.0J$ ($L=128$). $c)$ Steady-state density and $d)$ current profiles for a chain of length $L=4096$ at $\phi=\theta=\pi/2$ for different disorder strengths. The inset in $c)$ highlights the reduction of net boundary charge accumulation with increasing disorder and the one in $d)$ shows that the bulk current is independent of the system size. Other parameters: $\Gamma=\kappa=0.1J$.}
    \label{fig:steady_current_density_sizes}
\end{figure}

Here, we discuss the robustness of the results in the non-interacting limit ($\Delta=0$) against the presence of imperfections, namely coherent disorder and bulk dephasing.

First, we consider the stability against coherent disorder by incorporating an Anderson potential into the original model,
\begin{equation}
    \mathcal{H} = \mathcal{H}_0 + \sum_j \varepsilon_j c^\dagger_j c_j,
\end{equation}
where the on-site energies $\varepsilon_j$ are uniformly sampled from the interval $[-W/2, W/2]$. We first inspect the system under PBC and analyze the steady-state particle current as a function of dissipative phases $\phi$ and $\theta$ for a substantial disorder strength $W=J$ [Fig.~\ref{fig:steady_current_density_sizes}(a)] compared to the clean limit [Fig.~\ref{fig:steady_current_density_sizes}(b)]. Notably, the current profile remains qualitatively similar to the clean limit: the steady-state current is maximal in absolute value for $\phi=\theta=\pm\pi/2$ and preserves the same orientation as in the clean case. However, there is a decrease in magnitude compared to the clean limit, as expected in the presence of a disorder potential. The particle current still vanishes along the line $\phi=-\theta$; we analytically demonstrate that precisely along this line, the dynamics continue to drive the system to a fully mixed state, as we proof in detail in Sec.~\ref{sec:steady_state_critical_line}.
\begin{figure}[t]
    \centering
    \includegraphics{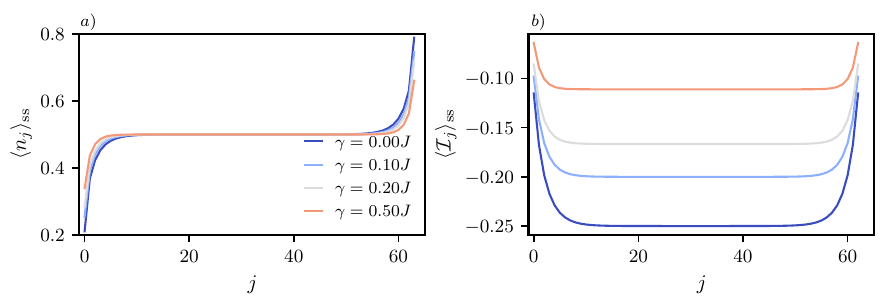}
    \caption{Steady-state particle density $a)$ and current $b)$ for increasing values of dephasing strength $\gamma$. Other parameters: $L=64$, $\Gamma=\kappa=0.1J$ and $\phi=\theta=\pi/2$.}
    \label{fig:current_density_dephasing}
\end{figure}
We now focus on the regime with OBC where, for $\phi=\theta=\pi/2$, the steady-state is non-reciprocal. In panels (c) and (d) of Fig.~\ref{fig:steady_current_density_sizes}, we show the steady-state density and current profiles, respectively, for a chain of length $L=4096$ at five distinct disorder strengths. Crucially, to rule out finite-size effects, we selected a chain length that exceeds the single-particle localization length of the Hamiltonian for all values of disorder considered. We observe that the qualitative features of the clean limit are preserved even when $W > J$. Specifically, the system exhibits charge accumulation (depletion) at the right (left) edges, alongside a finite bulk steady-state current. While increasing disorder tends to suppress the bulk current and reduce the charge imbalance at the edges, the current remains finite and the edge features remain distinct even at considerable disorder strengths, $W/J=2$ and $W/\Gamma=20$. Moreover, as shown in the inset of panel~(b), the steady-state current is independent of the system size for the lengths considered.

Finally, beyond static disorder, we also investigate the impact of local dephasing processes, as shown in Fig.~\ref{fig:current_density_dephasing}. We model this by coupling every site to a local Markovian bath with quantum jump operators $L_j = \sqrt{\gamma} c^\dagger_j c_j$. Once again, the effects of dephasing are qualitatively similar to those of static coherent disorder: the particle accumulation or depletion at the edges tends toward the bulk value [Fig.~\ref{fig:current_density_dephasing}.(a)], and the bulk current tends to be suppressed. We note, however, that the main observations discussed hold even for a dephasing strength considerably larger than the rates of the non-reciprocal pump and loss baths, such as $\gamma/\Gamma = \gamma/\kappa = 5$.

Thus, these results unambiguously demonstrate that the reported effects remain robust not only under perturbative imperfections but also in the presence of strong disorder ($W > J, \Gamma, \kappa$).

\section{Linblad Spectrum and Dissipative Gap}\label{sec:lindblad_spectrum}

\begin{figure}[b]
    \centering
    \includegraphics{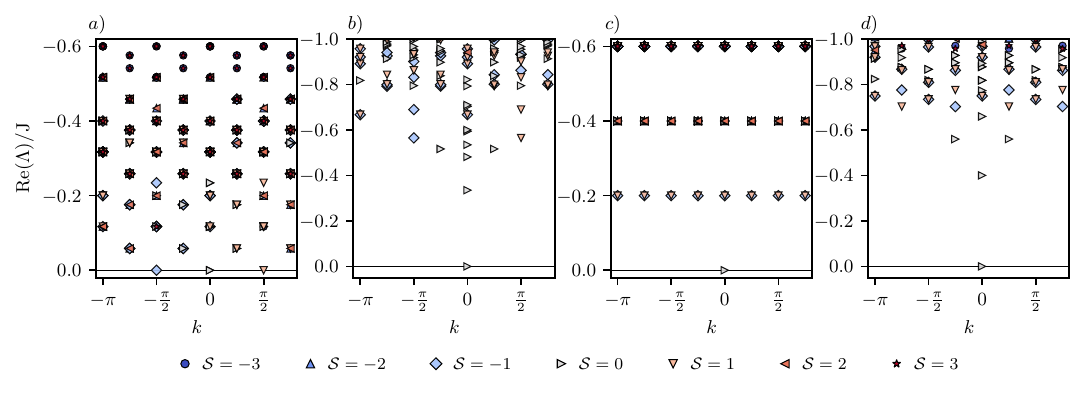}
    \caption{Real component of the Linblad spectra for $\phi=-\theta=\pi/2$ (panel $a)$ and $b)$) and for $\phi=\theta=\pi/2$ (panel $c)$ and $d)$). Panel $a)$ and $c)$ corresponds to the non-interacting limit, $\Delta=0.0J$, while $\Delta=4.0J$ in panel $c)$ and $d)$. Other parameters: $L=8$, $\Gamma=\kappa=0.1J$.}
    \label{fig:excitation_spectrum_symmetry}
\end{figure}
To better characterize the role of interactions, we analyze the spectrum of the Lindbladian superoperator. We use the super-fermion representation~\cite{Dzhioev2011,10.21468/SciPostPhys.9.6.091}, which maps the Lindbladian onto an augmented Hilbert space (the Liouville space), whose dimension with the number of sites $L$ scales as $4^L$. This construction doubles the degrees of freedom by introducing a \textit{tilde} space to encode the action on the bra of the density matrix (see Ref.~\cite{Dzhioev2011} for technical details). In this formalism the Linblad superoperator corresponds to,
\begin{equation}
\begin{aligned}
\mathcal{L}	&=\sum_{j}\left[J_{R}c_{j}^{\dagger}c_{j+1}+J_{L}c_{j+1}^{\dagger}c_{j}+J_{R}^{\ast}f_{j}^{\dagger}f_{j+1}+J_{L}^{\ast}f_{j+1}^{\dagger}f_{j}-\Delta\left(n_{j}n_{j+1}-\tilde{n}_{j}\tilde{n}_{j+1}\right)\right] \\
&+\left(\Gamma-\kappa\right)\sum_{j}\left[c_{j}^{\dagger}c_{j}+f_{j}^{\dagger}f_{j}\right]+i\kappa\sum_{j}\left[2f_{j}c_{j}+e^{i\phi}f_{j}c_{j+1}+e^{-i\phi}f_{j+1}c_{j}\right] \\
&+i\Gamma\sum_{j}\left[\left(2f_{j}^{\dagger}c_{j}^{\dagger}+e^{i\theta}f_{j}^{\dagger}c_{j+1}^{\dagger}+e^{-i\theta}f_{j+1}^{\dagger}c_{j}^{\dagger}\right)\right]-2L\Gamma.
\end{aligned}
\end{equation}
where $f_j(f^\dagger_j)$ are fermion creation(annihilation) operators that act on the \textit{tilde} Fock space that satisfy the standard fermionic anticommutation relations, $J_{R}=\frac{i}{2}J+\frac{1}{2}\kappa e^{i\phi}-\frac{1}{2}\Gamma e^{-i\theta}$ and $J_{L}=\frac{i}{2}J+\frac{1}{2}\kappa e^{-i\phi}-\frac{1}{2}\Gamma e^{i\theta}$.

We obtain the symmetry-resolved spectra of the Linblad by employing translational symmetries (when considering PBC) and the $U(1)$ symmetry generated by $\mathcal{S}=\sum_j \left( c^\dagger_jc_j - f^\dagger_jf_j \right)$ as $\left[\mathcal{L}, \mathcal{S} \right]=0$. The latter is a consequence of the weak $U(1)$ symmetry related to the particle number operator. We note that all Hermitian eigenmatrices of the Lindblad superoperator (which have real eigenvalues) belong to the \( \mathcal{S}=0 \) sector.

In the non-interacting limit, the vectorized Lindblad superoperator admits an exact analytic diagonalization, yielding the same single-particle spectrum as obtained from the retarded Keldysh Green’s function in section~\ref{sec:keldysh_section},
\begin{equation}
    \mathcal{L} = \left( -i \varepsilon_k - \frac{1}{2}\lambda_k \right) D^\dagger_k d_k +\left( i \varepsilon_k - \frac{1}{2}\lambda_k \right) \tilde{d}^\dagger_{-k} \tilde{D}_{-k},
    \label{eq:free_linblad}
\end{equation}
where $d_k, D_k^\dagger$ and $\tilde d_k, \tilde D_k^\dagger$ denote fermionic modes acting on the original and \textit{tilde} Fock spaces, respectively. Within this formalism, the steady state is defined as the state annihilated by all \( d_k \) and \( \tilde D_k \), while the decaying modes are generated by the action of \( D_k^\dagger \) and \( \tilde d_k^\dagger \) upon it.

\begin{figure}[t]
    \centering
    \includegraphics{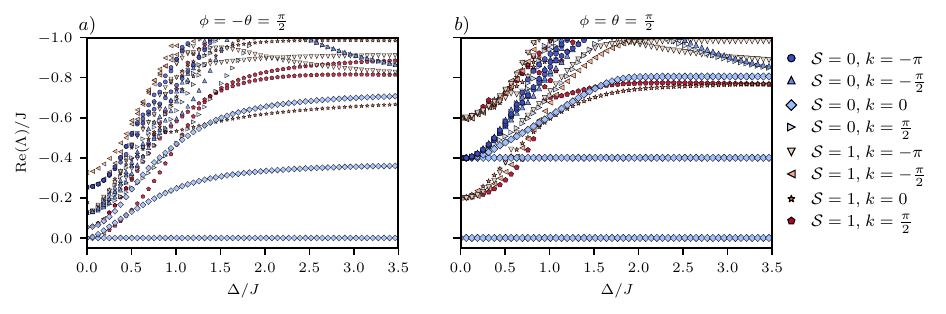}
    \caption{Real part of the spectrum as a function of $\Delta/J$ for $\phi=-\theta=\pi/2$ $a)$ and $\phi=\theta=\pi/2$ $b)$. Other parameters: $L=12$ and $\Gamma=\kappa=0.1J$.}
    \label{fig:symmetry_resolved_spectrum_interactions}
\end{figure}
We now discuss the structure of the real part of the Linblad spectra, which is responsible for the decaying of the modes. In Fig.~\ref{fig:excitation_spectrum_symmetry}, we plot the symmetry resolved real-part of the spectrum for different parameters. In the non-interacting limit defined by $\phi=-\theta$ (Fig.~\ref{fig:excitation_spectrum_symmetry}(a)), we observe four eigenvalues with zero real parts. Two of these correspond to Hermitian density matrices with zero momentum. These states represent the degenerate steady states:
\begin{equation}
    \rho^{(1)}_{\mathrm{ss}} = \dfrac{\mathbb{I}_{2^N\times 2^N}}{2^N}, \quad \rho^{(2)}_{\mathrm{ss}} = \dfrac{c^\dagger_{k^\ast}c_{k^\ast}}{2^{N-1}}.
\end{equation}
The remaining two states lie in the $\mathcal{S}=\pm1$ sectors with momenta $k^\ast$ and $-k^\ast$, respectively, corresponding to the non-normalized density matrices
\begin{equation}
    \rho_{\mathcal{S}=1,k=k^\ast} = c^\dagger_{k^\ast},\quad \rho_{\mathcal{S}=-1,k=-k^\ast} = c_{k^\ast}.
\end{equation}
This is a natural consequence of the dark mode of the Linblad as modes with momentum $k^\ast$ are transparent to the dissipation. In contrast, whenever $\phi\neq-\theta$, all modes acquire finite dissipation, as illustrated in Fig.~\ref{fig:excitation_spectrum_symmetry}(c) for the case $\phi=\theta=\pi/2$. Specifically, for these angles and with $\Gamma=\kappa$, the dissipative component becomes momentum-independent, given by $\lambda_k=4\Gamma$. This is reflected in a constant spacing of $2\Gamma$ observed between the different eigenvalues. In this regime, the steady state resides in the zero-momentum sector, while the second longest-lived modes belong to the $\mathcal{S}=\pm1$ sectors, corresponding to the operators:
\begin{equation}
    \rho_{\mathcal{S}=1,k} = c^\dagger_{k} \rho_{\mathrm{ss}}, \quad \rho_{\mathcal{S}=-1,k} = \rho_{\mathrm{ss}}c_{k}.
\end{equation}
We also observe that in this non-interacting limit, most eigenvalues are degenerate with respect to the real part. The degree of degeneracy naturally depends on the system parameters and can be predicted from Eq.~\eqref{eq:free_linblad}. On the other hand, in the presence of interactions, most of these degeneracies are lifted\footnote{ The remaining ones stem from the trace-preserving and Hermiticity-preserving nature of the Lindblad superoperator.} and there is the opening of the many-body dissipative gap for $\phi=-\theta=\pi/2$ Fig.~\ref{fig:excitation_spectrum_symmetry}(b) or its enhancement for $\phi=\theta=\pi/2$ as shown in Fig.~\ref{fig:excitation_spectrum_symmetry}(c).

To better characterize the dissipative gap with the interaction strength, in Fig.~\ref{fig:symmetry_resolved_spectrum_interactions}, we show the dependence of the symmetry-resolved decay spectra on the interaction strength for $\phi=-\theta=\pi/2$ (Fig.~\ref{fig:symmetry_resolved_spectrum_interactions} (a) and $\phi=\theta=\pi/2$ (Fig.~\ref{fig:symmetry_resolved_spectrum_interactions} (b). Owing to translational symmetry, the steady state remains in the zero-momentum sector with $\mathcal{S}=0$, which for the critical line $\phi=-\theta$, it is insensitive to the interaction term, $\Delta$, and it always corresponds to the infinite-temperature steady state just like in the non-interacting limit (check section~\ref{sec:steady_state_critical_line} for the proof),
\begin{equation}
\rho_{\rm ss,\phi=-\theta} = \bigotimes_{j=1}^{L} \left( \dfrac{\kappa}{\Gamma+\kappa}\left|0 \right\rangle \left\langle 0 \right| + \dfrac{\Gamma}{\Gamma+\kappa}\left|1 \right\rangle \left\langle 1 \right| \right),
\label{eq:steady_state_density_matrix}
\end{equation}
with a net particle density fixed by the pump and loss rates, $\langle n_j\rangle_{\rm ss} = \Gamma/(\Gamma+\kappa)$. From Fig.~\ref{fig:symmetry_resolved_spectrum_interactions} (a), we observe that the dissipative gap increases monotonically with increasing $\Delta$. In the weak-interaction regime $\Delta \ll J$, the slowest relaxation rate is determined by a mode in the $\mathcal{S}=\pm1$ sector and $k=\mp k^\ast$. By contrast, for $\Delta \sim J$ and larger, the slowest mode lies in the zero-momentum sector with $\mathcal{S}=0$ and can be continuously connected, in the non-interacting limit, to the degenerate steady state. Given the possible system sizes accessible within our computational resources, results obtained in the limit $\Delta \to 0$ require careful interpretation regarding finite-size corrections. A critical factor is the closure of the dissipative gap at a specific finite momentum. Consequently, if this momentum is incommensurate with the discrete momentum grid defined by the chain length, $k=2\pi \, n/L$, with $n\in\mathbb{Z}_L$, the true minimum of the dissipative spectrum may remain unresolved.

At $\phi=\theta=\pi/2$, the system exhibits a gap even in the non-interacting limit. However, interactions further enhance this dissipative gap. With increasing $\Delta$, the quantum numbers of the gap change. Initially determined by a finite-momentum state in the $\mathcal{S}=\pm 1$ sector, the gap becomes dictated by a zero-momentum state in the $\mathcal{S}=0$ sector for $\Delta > J$. Notably, the decay rate associated with this zero-momentum state is insensitive to interaction strength. This robustness arises because, at $\Gamma=\kappa$, the state corresponds to the infinite-temperature density matrix, which is insensitive to the Hamiltonian term.

\section{Steady-state density-matrix along the critical line $\phi=-\theta$}
\label{sec:steady_state_critical_line}
In this appendix, we prove that the density matrix given in Eq.~\eqref{eq:steady_state_density_matrix} corresponds to the steady-state density matrix along the line $\phi=-\theta$, independent of $\Delta$ and the boundary conditions.

First, we observe that Eq.~\eqref{eq:steady_state_density_matrix} can be explicitly written as a Gaussian density matrix,
\begin{equation}
    \rho_{\text{ss},\phi=-\theta} = \frac{1}{\mathcal{Z}} \exp\left(\ln\left(\dfrac{\Gamma}{\kappa}\right) N \right),
\end{equation}
where $\mathcal{Z}=\left(\kappa/(\Gamma+\kappa) \right)^L$ is the normalization factor and $N=\sum_{j} n_j$ is the total particle number operator. As the Hamiltonian conserves the total particle number (i.e., $\left[\mathcal{H},N \right]=0$) even in the presence of interactions (or the disorder potential considered in Sec.~\ref{sec:disorder_dephasing}) and for any boundary conditions, and because $\rho_{\text{ss}}$ is strictly a function of $N$, it follows that this density matrix does not evolve under the coherent part of the evolution:
\begin{equation}
    \left[\mathcal{H},\rho_{\text{ss},\phi=-\theta}\right] = 0.
\end{equation}

To prove that the density matrix is invariant under the dissipative part of the Lindblad master equation, it is useful to write the density matrix in the product state form:
\begin{equation}
    \rho_{\text{ss},\phi=-\theta} = \frac{1}{\mathcal{Z}}\bigotimes_{j=1}^L \tilde{\rho}_j,
    \label{eq:decompositon_inf_temp}
\end{equation}
with $\tilde{\rho}_j=\left(1+(\Gamma/\kappa-1)n_j \right)$. This exponential form yields the following useful commutation relation for any annihilation operator $c_m$:
\begin{equation} \label{eq:commutation_relation}
    c_{m}\;\rho_{\text{ss},\phi=-\theta} = \left(\frac{\Gamma}{\kappa}\right)\rho_{\text{ss},\phi=-\theta}\; c_{m}.
\end{equation}
The dissipative part that acts on $\rho_{\text{ss},\phi=-\theta}$ due to both gain and loss processes is given by
\begin{equation}
    \mathcal{D}[\rho_{\text{ss},\phi=-\theta}]=\sum_j\left( L_{j,\ell}\; \rho_{\text{ss},\phi=-\theta} \;L^\dagger_{j,\ell} + L_{jg} \;\rho_{\text{ss},\phi=-\theta}\; L^\dagger_{j,g}  -\frac{1}{2}\left\{L^\dagger_{j,\ell}L_{j,\ell}, \rho_{\text{ss},\phi=-\theta}   \right\} -\frac{1}{2}\left\{L^\dagger_{j,g}L_{j,g}, \rho_{\text{ss},\phi=-\theta}  \right\} \right).
\end{equation}
We note that $\left[\sum_j L^\dagger_{j,\mu} L_{j,\mu}, N \right]=0$ for $\mu \in \{\ell, g\}$. Consequently $\left[\sum_j L^\dagger_{j,\mu} L_{j,\mu}, \rho_{\text{ss},\phi=-\theta} \right]=0$, and so the anti-commutator terms simplify to a standard product. Using this last property and the commutation relation in Eq.~\eqref{eq:commutation_relation}, we can rewrite the dissipator as:
\begin{equation}
\begin{aligned}
\mathcal{D}[\rho_{\text{ss},\phi=-\theta}] &=\sum_j\left( \left[ \frac{\Gamma}{\kappa}L_{j,\ell} L^\dagger_{j,\ell} -L^\dagger_{j,g} L_{j,g} \right] \rho_{\text{ss},\phi=-\theta} + L_{j,g}\; \rho_{\text{ss},\phi=-\theta} \;L^\dagger_{j,g}  - \frac{\Gamma}{\kappa} L^\dagger_{j,g}\; \rho_{\text{ss},\phi=-\theta}\; L_{j,g}\right).
\end{aligned}
\end{equation}
Replacing the jump operators with their explicit expressions, we obtain:
\begin{equation}
\begin{aligned}
\mathcal{D}[\rho_{\text{ss},\phi=-\theta}] &= \Gamma\left( \left[e^{i\phi}-e^{-i\theta}\right]c_{j+1}c_{j}^{\dagger} + \left[e^{-i\phi}-e^{i\theta}\right]c_{j}c_{j+1}^{\dagger} \right)\rho_{\text{ss},\phi=-\theta} \\
&\quad + \left(e^{i\theta}-e^{-i\phi}\right)c_{j+1}^{\dagger}\rho_{\text{ss},\phi=-\theta} c_{j} + \left(e^{-i\theta}-e^{i\phi}\right)c_{j}^{\dagger}\rho_{\text{ss},\phi=-\theta} c_{j+1}.
\end{aligned}
\end{equation}
Finally, we observe that if $\phi=-\theta$, the prefactors in parentheses vanish. Thus, we conclude that:
\begin{equation}
    \phi=-\theta \implies \mathcal{D}[\rho_{\text{ss},\phi=-\theta}]=0.
\end{equation}
Given the decomposition in Eq.~\eqref{eq:decompositon_inf_temp}, it follows that all off-diagonal two-point correlators vanish. The only non-zero contribution arises from the diagonal terms corresponding to the local particle density:
\begin{equation}
    \langle c^\dagger_m c_n \rangle = \delta_{mn} \frac{\Gamma}{\Gamma+\kappa}.
\end{equation}
This result implies a vanishing particle current for every $\Delta$, which corroborates the numerical results presented in the main text.

\section{Lindbladian in the Rotating Frame and Doublon only Dynamics}

Here we briefly discuss the dynamics in the regime $\Delta\gg J$. To this extent it is convenient to perform a unitary transformation and move into a rotating frame, i.e., define a new density matrix $\Tilde{\rho}=V\rho V^{\dagger}$ with $V$ unitary operator of the form
\begin{align}
V=\exp(-iKt),
\end{align}
and $K$ a suitable operator. In this frame the Lindblad equation for the density matrix  $\Tilde{\rho}$ reads
\begin{equation}\label{eqn:lindblad2}
\partial_{t}\Tilde{\rho}=-i\left[\Tilde{\mathcal{H}},\Tilde{\rho}\right]+\sum_{j\mu}\Tilde{L}_{j\mu}\Tilde{\rho} \Tilde{L}_{j\mu}^{\dagger}-\dfrac{1}{2}\left\{ \Tilde{L}_{j\mu}^{\dagger}\Tilde{L}_{j\mu},\Tilde{\rho}\right\},
\end{equation}
where we have defined respectively the rotated Hamiltonian $\Tilde{\mathcal{H}}$
\begin{align}
\Tilde{\mathcal{H}}=i\left(\partial_t V\right)V^{\dagger}+V HV^{\dagger},    
\end{align}
and the rotated jump operators
\begin{align}
\Tilde{L}_{j\mu}=V L_{j\mu}V^{\dagger},\,\qquad\,
\Tilde{L}^{\dagger}_{j\mu}=V L^{\dagger}_{j\mu}V^{\dagger}.
\end{align}
We choose the unitary as 
\begin{align}
V=\exp(-i\Delta D t),
\end{align}
where $D=\sum_{i} n_in_{i+1}$ is the total number of next-neighbor doublons. This is picked, so that, the rotated Hamiltonian does not contain anymore the interaction term~\cite{seetharam2018absence,peronaci2018resonant}, namely
\begin{equation}
\Tilde{\mathcal{H}}=i\left(\partial_t V\right)V^{\dagger}+V \mathcal{H}V^{\dagger}=V \mathcal{H}_0V^{\dagger} 
\end{equation}
with $\mathcal{H}_0=-\frac{J}{2}\sum_{j=0}^{L-2}\left[c_{j}^{\dagger}c_{j+1}+c_{j+1}^{\dagger}c_{j}\right]$. The price to pay is the rotation of single particle operators in the frame generated by $V$. After simple manipulation these take the form
\begin{align}
Vc_jV^{\dagger}&= c_j\,e^{i\Delta (n_{j-1}+n_{j+1})t}.
\end{align}
We see from this expression that the dressed destruction operator now gains a phase factor that depends on the occupation of neighboring sites. In particular, starting from an occupied site $i$ we see that if either $j-1$ or $j+1$ are occupied, that is, in the presence of a doublon, the destruction operator gets a phase that rapidly oscillates at large $\Delta$ and thus averages out. Similarly for the creation operator we get
\begin{equation}
Vc^{\dagger}_jV^{\dagger}=c^{\dagger}_j\,e^{-i\Delta (n_{j-1}+n_{j+1})t}.
\end{equation}
As a consequence the hopping term entering the renormalized Hamiltonian reads
\begin{align}
\Tilde{\mathcal{H}}&= V\mathcal{H}_0V^{\dagger} = \sum_j e^{-i\Delta t(n_{j-1}-n_{j+2})}c^{\dagger}_jc_{j+1}+\text{h.c.},
\end{align}    
For what concerns the dissipative processes, we see that the dressed loss term corresponds to
\begin{align}
\Tilde{L}_{j,\ell}=\sqrt{\kappa}(e^{i\Delta (n_{j-1}+n_{j+1})t} c_j+ e^{i\phi}e^{i\Delta t(n_{j}+n_{j+2})t} c_{j+1}),
\end{align}
while the gain reads
\begin{align}
\Tilde{L}_{j,g}=\sqrt{\Gamma}(e^{-i\Delta (n_{j-1}+n_{j+1})t} c^{\dagger}_j+ e^{i\theta}e^{-i\Delta t(n_{j}+n_{j+2})t} c^{\dagger}_{j+1}).
\end{align}
In the rotated frame, the dynamics is described by a correlated hopping process with correlated gain and losses. In particular, coherent evolution contains resonant hopping processes between sites $j,j+1$ that conserve the doublon number (if $n_{j-1}=n_{j+2}$) and off-resonant ones, for $n_{j-1}\neq n_{j+2}$, which rapidly oscillate at frequency $\Delta$ and average out for $\Delta\gg J$ leading to doublon-only dynamics. In this regime, both gain and loss processes are also kinetically blocked, as they can only connect states that differ in doublon number and thus dissipate with a rapidly oscillating phase. As a result, for $\Delta \gg J$, the dynamics become increasingly coherent, and charge accumulation at the boundary is suppressed, as we observe numerically.

\section{Methods: Quantum Trajectories and Faber Polynomials}
\label{appendix:quantum_traj}

To solve both the dynamics and the steady-state correlation functions of the interacting model, we unravel the Lindblad master equation into quantum trajectories~\cite{daley2014quantum}. This approach is numerically more efficient, as it avoids evolving the full density matrix by instead simulating a smaller set of states from which the Lindblad dynamics are recovered through averaging.

In the following, we describe the quantum jump dynamics of the many-body Lindbladian in Eq.~(2) of the main text by unravelling the master equation using quantum jump trajectories. Specifically, we introduce the pure state $\vert \psi(\xi_t,t)\rangle$ which evolves according to the stochastic Schr\"odinger equation,
\begin{equation}
\begin{aligned}
d\ket{\psi_{\xi_t}(t )} 
&= -i dt \left[\mathcal{H}-\frac{i}{2}\sum_{j\mu} \left(L^\dagger_{j\mu} L_{j\mu}-\langle L^\dagger_{j\mu} L_{j\mu}\rangle_t \right)\right]\ket{\psi_{\xi_t}(t )} 
+ \sum_{j,\mu}\left(\frac{L_{j,\mu}}{\sqrt{\langle L^\dagger_{j,\mu} L_{j,\mu}\rangle}}-1\right) d\xi_{j,\mu,t}\ket{\psi(\xi_t,t )}, 
\end{aligned}
\label{eq:qjump}
\end{equation}
where $\langle\circ\rangle_t\equiv \langle \psi(\xi_t,t) \vert\circ\vert \psi(\xi_t,t)\rangle$, and $\xi_t=\left\{\xi_{j,\mu,t}\right\}$ represent a collection of independent Poisson processes ${d\xi_{j,\mu,t}\in\{0,1\}}$ characterized by the mean $\overline{d\xi_{j,\mu,t}} = dt \langle L^\dagger_{j,\mu} L_{j,\mu}\rangle_t$.This dynamic arises from a sequence of random quantum jumps, causing sudden changes in the wave function, alongside a deterministic, non-unitary evolution governed by a non-Hermitian Hamiltonian
\begin{equation}
    \mathcal{H}_{\rm nH}=\mathcal{H}-\frac{i}{2}\sum_{j,\mu} L^\dagger_{j,\mu} L_{j,\mu}.
\end{equation}
If one considers the average of the pure state density matrix over all quantum jump trajectories, the resulting density matrix takes the form i.e.
\begin{equation}
    \rho(t)=\overline{\vert \psi(\xi_t,t)\rangle\langle  \psi(\xi_t,t)\vert},
\end{equation}
which evolves according to the Linblad master equation written in Eq.~(2) of the main text. In order to perform the non-Hermitian evolution between consecutive quantum jumps, we make use of the recently developed Faber polynomial method~\cite{Soares2024Faber}. 

The dynamics of each quantum trajectory is numerically simulated by combining the high-order Monte Carlo Wave Function algorithm~\cite{daley2014quantum,Landi2024_prx_quantum} with the Faber polynomial algorithm presented in~\cite{Soares2024Faber}. In particular, between quantum jumps, the state's non-unitary evolution is expanded using a Faber polynomial series expansion of the time evolution operator
\begin{equation}
e^{-i\mathcal{H_{\rm nH}}t}\ket{\Psi \left(t_0\right)}=\sum_{n=0}^{+\infty}c_n \left(t \right)F_{n}\left(\dfrac{\mathcal{H_{\rm nH}}}{\lambda}\right) \ket{\Psi \left(t_0\right)}, \quad c_n \left(t \right)=e^{-i\lambda\gamma_{0}t}\left(\dfrac{-i}{\sqrt{\gamma_{1}}}\right)^{n}J_{n}\left(2\sqrt{\gamma_{1}}\lambda t\right),
\label{eq:Faber_expansion}
\end{equation}
where $F_n \left( \cdot \right)$ is the $\rm n^{th}$ Faber Polynomial, $J_n \left( \cdot \right)$ is the $\rm n^{th}$ Bessel function of the first kind, $\lambda$ is a rescaling parameter and $\gamma_0$ and $\gamma_1$ are related with the bounds of the spectrum of the non-Hermitian Hamiltonian (for further details check consult~\cite{Soares2024Faber}). We never explicitly compute the $F_n \left(\mathcal{H}_{\rm nH}/\lambda \right)$, but rather do it implicit when acting on the states using the recurrence relation satisfied by the polynomials,
\begin{equation}
\begin{aligned}
\ket{\Psi_{0}} & =\ket{\Psi\left(t_{0}\right)},\\
\ket{\Psi_{1}} & =\left(\dfrac{\mathcal{H}_{\rm nH}}{\lambda}-\gamma_{0}\right)\ket{\Psi_{0}},\\
\ket{\Psi_{2}} & =\left(\dfrac{\mathcal{H}_{\rm nH}}{\lambda}-\gamma_{0}\right)\ket{\Psi_{1}}-2\gamma_{1}\ket{\Psi_{0}},\\
\ket{\Psi_{n+1}} & =\left(\dfrac{\mathcal{H}_{\rm nH}}{\lambda}-\gamma_{0}\right)\ket{\Psi_{n}}-\gamma_{1}\ket{\Psi_{n-1}},\quad n > 2,
\end{aligned}
\label{eq:req_faber}
\end{equation}
where $\ket{\Psi_n} = F_n \left(\mathcal{H}_{\rm nH}/\lambda \right) \ket{\Psi\left(t_0\right)}$. As discussed in Ref.~\cite{Soares2024Faber}, the computational demand of comes from acting with the operation of multiplying a state by the Hamiltonian matrix, which has a linear cost with the Hilbert space dimension (as long as the Hamiltonian matrix is sparce on some basis). Moreover, the memory cost also scales linearly as only two vectors, $\ket{\Psi_{n-1}}$ and $\ket{\Psi_n}$, must be saved in memory.

\bibliography{refs}